\pdfoutput=1

\documentclass[aps,showpacs,twocolumn,twoside,prc,superscriptaddress]{revtex4-1}
\usepackage{epsfig,amsmath,amssymb}
\usepackage{graphicx}
\usepackage{array}

\def\I#1{\!#1\!}
\def\vecb#1{\boldsymbol{#1}}
\def\ket#1{|#1\rangle}

\def\matr#1#2#3{\langle#1|#2|#3\rangle}
\def\abs#1{\left\lvert#1\right\rvert}

\def\ave#1{\langle#1\rangle}

\def\={\!=\!}
\def\>{\!>\!}
\def\<{\!<\!}
\def\-{\!-\!}
\def\+{\!+\!}

\def\abs#1{\left|#1\right|}

\def\uvo#1{\lq\lq #1\rq\rq}

\def\bz{\beta_0'}

\def\su3{\mathrm{SU3}}
\def\u5{\mathrm{U5}}

\newcommand{\ba}{\begin{eqnarray}}
\newcommand{\ea}{\end{eqnarray}}
\newcommand{\bmath}{\begin{mathletters}}
\newcommand{\emath}{\end{mathletters}}
\newcommand{\ban}{\begin{eqnarray*}}
\newcommand{\ean}{\end{eqnarray*}}

\newcommand{\bsub}{\begin{subequations}}
\newcommand{\esub}{\end{subequations}}

\begin{document}

\title{Excited-state quantum phase transitions in systems with two degrees of freedom: 
\\
III. Interacting boson systems}

\author{Michal Macek}
\affiliation{The Czech Academy of Sciences, Institute of Scientific Instruments, Kr\'alovopolsk\'a 147, 61264 Brno, Czech Republic}
\author{Pavel Str\'ansk\'y}
\affiliation{Institute of Particle and Nuclear Physics, Faculty of Mathematics and Physics, Charles University, V Hole\v{s}ovi\v{c}k\'ach 2, 18000 Prague, Czech Republic} 
\author{Amiram Leviatan}
\affiliation{Racah Institute of Physics, The Hebrew University, Jerusalem 91904, Israel}
\author{Pavel Cejnar}
\affiliation{Institute of Particle and Nuclear Physics, Faculty of Mathematics and Physics, Charles University, V Hole\v{s}ovi\v{c}k\'ach 2, 18000 Prague, Czech Republic} 

\date{\today}

\begin{abstract}
The series of articles [Ann.\,Phys. 345, 73 (2014)  and 356, 57 (2015)] devoted to excited-state quantum phase transitions (ESQPTs) in systems with ${f=2}$ degrees of freedom is continued by studying the interacting boson model of nuclear collective dynamics as an example of a truly many-body system.
The intrinsic Hamiltonian formalism with angular momentum fixed to ${L=0}$ is used to produce a generic first-order ground-state quantum phase transition with an adjustable energy barrier between the competing equilibrium configurations. 
The associated ESQPTs are shown to result from various classical stationary points of the model Hamiltonian, whose analysis is more complex than in previous cases because of (i) a non-trivial decomposition to kinetic and potential energy terms and (ii) the boundedness of the associated classical phase space.
Finite-size effects resulting from a partial separability of both degrees of freedom are analyzed.
The features studied here are inherent in a great majority of interacting boson systems.
\end{abstract}


\maketitle

\section{Introduction}
\label{sec:Intro}

Excited-state quantum phase transitions (ESQPTs) \cite{ref:Cejn06,ref:Capr08,ref:Cejn08,ref:Cejn09,ref:E1,ref:E2,ref:Stra16} represent an extension of the ground-state quantum phase transitions (QPTs) \cite{ref:Carr11} to the eigenstates with higher energy.
ESQPTs were introduced as robust singularities in the spectra of energy levels in the nuclear interacting boson model (IBM) \cite{ref:Cejn06,ref:Hein06,ref:Mace06}, and have since been studied theoretically in numerous other many-body models like the Lipkin model, see, e.g., Refs.\,\cite{ref:Ribe07,ref:Opat18}, the molecular vibron model~\cite{ref:Lare11,ref:Lare13}, the fermionic and bosonic pairing models \cite{ref:Skra11,ref:Rela16}, the extended Dicke model of superradiance \cite{ref:Bran13,ref:Basta14,ref:Kloc17a,ref:Kloc17b}, the Bose-Hubbard model of atom-molecule condensates \cite{ref:Rela14,ref:Bych18}, and in algebraic models of two-dimensional crystals \cite{ref:Diet13,ref:Iach15,ref:Diet17}.
The ESQPTs were experimentally observed in molecules~\cite{ref:Lare11,ref:Lare13} and in some artificial quantum systems like photonic crystals \cite{ref:Diet13}.
Dynamical consequences of ESQPTs in the response to various kinds of driving \cite{ref:Ferna11,ref:Basti14,ref:Pueb15,ref:Kopy15,ref:Sant16,ref:Kopy17,ref:Kloc18} or in the entanglement and decoherence properties \cite{ref:Rela08,ref:Kloc17a} are of potential relevance for quantum simulation and computation applications.
Also the thermodynamic anomalies related to ESQPTs \cite{ref:Basta16,ref:Fern16,ref:Cejn17} are of fundamental interest. 

For a general bound quantum system with quantum Hamiltonian $\hat{H}^\lambda$ depending on a control parameter $\lambda$, the ESQPT represents a singularity in the average density and slope of discrete energy levels $E_i^{\lambda}$ and in the overall structure of the corresponding eigenstates $\ket{\psi_i^{\lambda}}$ (in this paper, the dependence on the control parameter is always expressed by the superscript $\lambda$, which should not be confused with the power symbol).
The singularity is caused by a quasi-stationary point of the underlying classical dynamics. 
A general classification of such singularities for non-degenerate stationary points in systems with an arbitrary number of degrees of freedom $f$ was given in Ref.\,\cite{ref:Stra16}.
A systematic analysis of ESQPTs in systems with $f=2$ was initiated in Refs.\,\cite{ref:E1} and \cite{ref:E2}.
In these articles, further referred as part~I and part~II, respectively, the focus was set to systems whose classical limit is expressed via coordinates $\vecb{q}$ and momenta $\vecb{p}$  in an unbounded phase space in which the Hamiltonian function ${\cal H}^{\lambda}(\vecb{q},\vecb{p})$ can be cast as a sum of a simple kinetic energy term ${\cal K}(\vecb{p})\I{\propto}\vecb{p}^2$ and an arbitrary potential ${\cal V}^{\lambda}(\vecb{q})$.
However, in the classical limit of quantum many-body systems the phase space is usually bounded and the Hamiltonian decomposition is more complicated.
In such systems, the ESQPT analysis shows some peculiarities.

The present article (part III) continues the series of Refs.\,\cite{ref:E1,ref:E2} by considering a suitable example of a many-body system with $f=2$, namely the IBM~\cite{ref:IBMbook} with angular momentum set to $L=0$.
The transition between the U(5) and O(6) dynamical symmetries of this model, which is connected with the second-order ground-state QPT, was at the beginning of the ESQPT development \cite{ref:Cejn06,ref:Hein06,ref:Mace06}.
More recently, ESQPTs along the transition between the U(5) and SU(3) dynamical symmetries with the first-order ground-state QPT were studied for the $L=0$ subset of the IBM spectrum \cite{ref:Zhan16}, as well as with respect to states with $L>0$ \cite{ref:Zhan17}.
Our present article also deals with the ESQPT structures accompanying the first-order QPT between spherical and axially deformed ground-state configurations.
The focus on a first-order QPT Hamiltonian is in agreement with the general strategy of parts I and II.
However, the setup of the model is different than in Refs.\,\cite{ref:Zhan16,ref:Zhan17}, that employed a simple \lq\lq consistent-$Q$\rq\rq\ IBM Hamiltonian.
Instead we apply the formalism introduced in Refs.\,\cite{ref:Kirs85,ref:Levi87} in the context of resolution of the IBM dynamics into intrinsic and collective components.
This approach turned out to be very useful in the study of first-order QPTs \cite{ref:Levi06,ref:Mace11,ref:Levi12,ref:Mace14} and we use here its advantage of providing an adjustable (preferably high) energy barrier between the competing ground-state configurations. 
This facilitates our study of typical ESQPT effects associated with generic first-order QPTs in quantum many-body systems. 

The article is organized as follows: 
In Sec.\,\ref{sec:MB}, we consider a general interacting boson system, construct its classical limit, and describe new general aspects of the ESQPT analysis that follow from the compactness of the relevant phase space and from non-separability of coordinate and momentum Hamiltonian terms.
In Sec.\,\ref{sec:IBM}, as a case study demonstrating general effects introduced in Sec.\,\ref{sec:MB}, we present a numerical analysis of ESQPTs in a family of IBM Hamiltonians describing the spherical--deformed nuclear shape-phase transition. 
In Sec.\,\ref{sec:Sum} we give a brief summary and outlook.

\section{General interacting boson systems}
\label{sec:MB}

\subsection{Hamiltonian and classical limit}
\label{sec:MBHa}

This article deals with systems of interacting bosons.
It may be extended to systems of interacting fermions by identifying a particular fermionic collective algebra and realizing it in terms of bosonic operators. 
This is usually possible, but not discussed here.
We further assume that the total number of particles (bosons) $N$ is conserved, so that it can naturally play the role of the size parameter $\aleph$ used in parts~I and II \cite{ref:E1,ref:E2}, hence we set $\aleph\equiv N$. 

In the following, we consider bosonic Hamiltonians with one- and two-body terms, namely 
\begin{equation}
\hat{H}^{\lambda}=\sum_{k,k'=0}^{n}\varepsilon^{\lambda}_{kk'}\ \hat{b}_{k}^{\dag}\hat{b}_{k'}+\frac{1}{N\I{-}1}\!\!\sum_{\left.\smallmatrix k,k'\\l',l\endsmallmatrix\right\}=0}^{n}\!\!\!\!v^{\lambda}_{kk'll'}\ \hat{b}_k^{\dag}\hat{b}_{k'}^{\dag}\hat{b}_{l'}\hat{b}_l
\label{Habo}
\,,
\end{equation}
where $\hat{b}_k^{\dag}$ and $\hat{b}_k$ are creation and annihilation operators of bosons in $n\I{+}1$ single-particle states enumerated by $k=0,1,2,\dots,n$ (with $n\geq 1$).
The $(n\I{+}1)^2$ bilinear products of boson creation and annihilation operators, or more precisely symmetric and antisymmetric Hermitian operators of the form $\hat{b}_k^{\dag}\hat{b}_l+\hat{b}_l^{\dag}\hat{b}_k$ and $i\bigl(\hat{b}_k^{\dag}\hat{b}_l-\hat{b}_l^{\dag}\hat{b}_k\bigr)$, represent generators of the system's dynamical algebra U($n\I{+}1$).
Coefficients $\varepsilon^{\lambda}_{kk'}$ and $v^{\lambda}_{kk'll'}$ in Eq.\,(\ref{Habo}) are assumed to depend on an adjustable control parameter $\lambda$ and must always satisfy the constraints following from the Hermiticity requirement.
The $N\I{-}1$ denominator ensures that the proportion between the one- and two-body terms is preserved for varying total number of bosons.

The Hamiltonian (\ref{Habo}) can be cast in the coordinate--momentum form with the aid of transformation 
\begin{equation}
\hat{b}_k^{\dag}=\sqrt{\frac{N}{2}}(\hat{q}_k-i\hat{p}_k),\quad
\hat{b}_k=\sqrt{\frac{N}{2}}(\hat{q}_k+i\hat{p}_k),
\label{como}
\end{equation}
where $\{\hat{q}_k,\hat{p}_k\}_{k=0}^{n}$ are operators of canonically conjugate coordinates and momenta.
The commutator
\begin{equation}
\bigl[\hat{q}_k,\hat{p}_l\bigr]=\frac{i}{N}\delta_{kl}
\label{comu}
\end{equation}
indicates that $N^{-1}$ represents an effective Planck constant of the system and $N\to\infty$ corresponds to its classical limit.
The classical Hamiltonian ${\cal H}^{\lambda}$ associated with $\hat{H}^{\lambda}$ in Eq.\,(\ref{Habo}) is obtained via the mapping 
\begin{equation}
\lim_{N\to\infty}\frac{\hat{H}^{\lambda}}{N}\ \mapsto\ {\cal H}^{\lambda}\bigl(\{\hat{q}_k,\hat{p}_k\}\bigr)
\,,
\label{Hcla}
\end{equation}
where the substitution (\ref{como}) is made in $\hat{H}^{\lambda}$ with the confidence that for ${N\to\infty}$ all the $\hat{q}_k$ and $\hat{p}_l$ terms mutually commute and become just ordinary numbers $q_k$ and $p_l$. 
These variables constitute a classical phase space, which is finite due to the conservation of ${\hat{N}=\sum_{k=0}^n\hat{b}_k^{\dag}\hat{b}_k}$.
Indeed, the relation expressing this constraint, ${\sum_{k=0}^n\bigl(q_k^2+p_k^2\bigr)=2}$, implies that this kind of phase space coincides with the surface of a $2(n\I{+}1)$-dimensional sphere with radius $\sqrt{2}$. 
But this is not yet the space applied in most of calculations.

One path to the proper classical limit of Hamiltonian \eqref{Habo} leads via a generalization of the Holstein-Primakoff transformation \cite{ref:Blai78}. 
The constraint on conserved observable $\hat{N}$ is used for a complete elimination of one degree of freedom,
e.g. that associated with boson $b_0$.
In particular, bosons $b_k$ are transformed to new bosons $a_k$,
\begin{equation}
\hat{a}_0^{\dag}=\sqrt{\frac{\hat{N}}{\hat{b}_0^{\dag}\hat{b}_0}}\ \hat{b}_0^{\dag}\,\quad
\hat{a}_k^{\dag}=\hat{b}_0\sqrt{\frac{1}{\hat{b}_0^{\dag}\hat{b}_0}}\ \hat{b}_k^{\dag}\  ({\rm for}\ k\I{>}0),
\label{trab}
\end{equation}
with the inverse transformation given by:
\begin{equation}
\hat{b}_0^{\dag}\I{=}\sqrt{\frac{\hat{a}_0^{\dag}\hat{a}_0-\sum\limits_{l=1}^{n}\hat{a}_l^{\dag}\hat{a}_l}{\hat{a}_0^{\dag}\hat{a}_0}}\hat{a}_0^{\dag},\ \
\hat{b}_k^{\dag}\I{=}\sqrt{\frac{1}{\hat{a}_0^{\dag}\hat{a}_0}}\hat{a}_0^{\dag}\hat{a}_k^{\dag}\  (k\I{>}0)
\label{trabi}
\end{equation}
(note that the operator denominators in these expressions do not produce singularities).
Since the $a_k$ operators satisfy bosonic commutation relations, the transformation is canonical.
However, the total number of the new bosons does not match with $\hat{N}$ and is not conserved by Hamiltonian (\ref{Habo}).
Instead we have ${\hat{a}_0^{\dag}\hat{a}_0=\hat{N}}$ and ${\sum_{l=1}^{n}\hat{a}_l^{\dag}\hat{a}_l=\hat{N}-\hat{b}_0^{\dag}\hat{b}_0}$.
This means that the conserved number $\hat{N}$ is \lq\lq stored\rq\rq\ solely in boson $a_0$ while the active degrees of freedom are represented by the remaining bosons  $a_k$ with ${k=1,...,n}$.
Hence we end up with a system with ${f=n}$ degrees of freedom.

Applying the coordinate--momentum decomposition (\ref{como}) to all $a_k$ bosons from Eq.\,(\ref{trab}), we obtain the classical-limit expressions of generators of the dynamical group ${{\rm U}(n+1)}$ in terms of coordinates and momenta ${\{q_k,p_k\}_{k=1}^n\equiv\{\vecb{q},\vecb{p}\}}$ associated with bosons $\{a_k\}_{k=1}^{k}$.
These expressions are given in Table~\ref{gene}.
We see that the classical representation is the same as if we applied the decomposition (\ref{como}) to the original bosons $b_k$ with a substitution ${\hat{b}_0^{\dag},\hat{b}_0\mapsto\sqrt{2-\sum_{l=1}^f(q_l^2+p_l^2)}}$.
Note that the new coordinates and momenta must satisfy the condition:
\begin{equation}
R^2\equiv\sum_{k=1}^n\bigl(q_k^2+p_k^2\bigr)\leq 2\equiv R_0^2\,,
\label{sphere}
\end{equation}
which means that the phase space with ${f=n}$ is the interior of a $2f$-dimensional sphere with radius ${R_0=\sqrt{2}}$.
Hereafter this domain is denoted by symbol ${\bf\Omega}$.

\begin{table}
\begin{ruledtabular}
\begin{tabular}{cccc}
          &          &   symmetric generators                                                                                        &  antisymmetric generators 
\\
$k$    & $l$     &   $\hat{b}_k^{\dag}\hat{b}_l+\hat{b}_l^{\dag}\hat{b}_k$  &  $i\bigl(\hat{b}_k^{\dag}\hat{b}_l-\hat{b}_l^{\dag}\hat{b}_k\bigr)$ 
\\
\hline
\vspace{-2mm}
\\
$>\!0$ & $>\!0$ &  $N\bigl(q_kq_l+p_kp_l\bigr)$ & $N\bigl(p_kq_l-q_kp_l\bigr)$ 
\\ 
$>\!0$ & $=\!0$ &  $Nq_k\sqrt{2\I{-}\!\sum\limits_{m=1}^n\bigl(q_m^2\I{+}p_m^2\bigr)}$ & $Np_k\sqrt{2\I{-}\!\sum\limits_{m=1}^n\bigl(q_m^2\I{+}p_m^2\bigr)}$  
\\ 
$=\!0$ & $>\!0$ &  $Nq_l\sqrt{2\I{-}\!\sum\limits_{m=1}^n\bigl(q_m^2\I{+}p_m^2\bigr)}$ & $-Np_l\sqrt{2\I{-}\!\sum\limits_{m=1}^n\bigl(q_m^2\I{+}p_m^2\bigr)}$  
\\ 
$=\!0$ & $=\!0$ &  $N\biggl(2\I{-}\!\sum\limits_{m=1}^n\bigl(q_m^2\I{+}p_m^2\bigr)\biggr)$ & $0$  
\end{tabular}
\end{ruledtabular}
\caption{Classical limit of the U($n\I{+}1$) generators.}
\label{gene}
\end{table}

Decomposing an arbitrary quantum Hamiltonian (\ref{Habo}) into the Hermitian generators of ${{\rm U}(n+1)}$, we can use the expressions in Table~\ref{gene} to write down its classical counterpart.
A common alternative to this construction is based on the use of various kinds of coherent states \cite{ref:Zhan90,ref:Iach14}, which finally does the same job.
In this case, the classical Hamiltonian ${\cal H}^{\lambda}$ is obtained from the energy-per-boson expectation value 
\begin{equation}
{\cal H}^{\lambda}(\alpha)=\frac{\matr{\alpha}{\hat{H}^{\lambda}}{\alpha}}{\matr{\alpha}{\hat{N}}{\alpha}}
\label{Haco}
\,.
\end{equation}
in a coherent state $\ket{\alpha}$, with ${\alpha\equiv\{\alpha_k\}_{k=0}^{n}}$ denoting a set of complex parameters composed of real coordinate and momentum variables $q_k$ and $p_k$.
In particular, one can use the Glauber coherent states given by:
\begin{equation}
\ket{\alpha}=e^{-\frac{\lVert\alpha\rVert^2}{2}} e^{\lVert\alpha\rVert\hat{B}^{\dag}(\tilde{\alpha})}\ket{0},\quad
\hat{B}^{\dag}(\tilde{\alpha})=\sum_{k=0}^n\tilde{\alpha}_k\hat{b}_k^{\dag}.
\label{co}
\end{equation}
Here $\ket{0}$ represents the boson vacuum and $\hat{B}^{\dag}(\tilde{\alpha})\ket{0}$ is a normalized single-boson state, which is in the basis $\{\hat{b}_k\ket{0}\}_{k=0}^{n}$ expressed by expansion coefficients ${\{\tilde{\alpha}_k\equiv\alpha_k/\lVert\alpha\rVert\}}$ with the normalization denominator ${\lVert\alpha\rVert=\sqrt{\sum_{k=0}^n|\alpha_k|^2}}$.
For the state \eqref{co}, the average number of bosons is ${\langle N\rangle_{\alpha}\I{\equiv}\matr{\alpha}{\hat{N}}{\alpha}=\lVert\alpha\rVert^2}$. 
Substituting ${\alpha_k=\sqrt{\frac{\langle N\rangle_{\alpha}}{2}}(q_k+ip_k)}$ for parameters with ${k=1,...,n}$ and setting ${\alpha_0=\sqrt{\langle N\rangle_{\alpha}-\sum_{l=1}^n|\alpha_k|^2}}$, one arrives at the same expression as obtained by the above procedure with $N$ replaced by $\langle N\rangle_{\alpha}$.

The projective coherent states,
\begin{equation}
\ket{\tilde{\alpha},N}=\frac{1}{\sqrt{N!}}\hat{B}^{\dag}(\tilde{\alpha})^N\ket{0},
\label{proco}
\end{equation}
 i.e., condensates  of a fixed number $N$ of bosons in the state $\hat{B}^{\dag}(\tilde{\alpha})\ket{0}$, yield equivalent results with
\begin{equation}
{\cal H}^{\lambda}(\tilde{\alpha},N)=\frac{1}{N}\matr{\tilde{\alpha},N}{\hat{H}^{\lambda}}{\tilde{\alpha},N}.
\label{Haproco}
\end{equation}
Note that the same can be achieved also with the other standard forms of coherent states (so-called algebraic and group coherent states) associated with the factorization ${{\rm U}(n+1)/[{\rm U}(n)\otimes{\rm U}(1)]}$ of the dynamical algebra \cite{ref:Iach14}.

The classical limit derived above holds for all boson systems that conserve $\hat{N}$.
Additional conservation laws may further reduce the number of effective degrees of freedom.
In this article series we investigate systems with ${f=2}$.
The bosonic model discussed in Sec.\,\ref{sec:IBM} has ${n=5}$, which corresponds to the full number of the model's degrees of freedom.
However, the invariance of the Hamiltonian under rotations in the 3-dimensional space makes it possible to fix the angular momentum to ${L=0}$, which reduces the number of active degrees of freedom by three.

\subsection{ESQPT analysis}
\label{sec:MBQu}

\subsubsection{ESQPTs due to non-degenerate stationary points}
\label{sec:cla}

Consider a general bound quantum system with Hamiltonian $\hat{H}^{\lambda}$ and discrete energy spectrum $E_i^{\lambda}$ enumerated by integer index $i$ and depending on a real control parameter $\lambda$. 
As discussed in parts~I and II \cite{ref:E1,ref:E2}, the ESQPT analysis relies on a decomposition $\bullet=\overline{\bullet}+\widetilde{\bullet}$ of the relevant quantities into smooth and oscillatory components.
In particular, we define the smoothed level density
\begin{equation}
\overline{\rho}^{\lambda}(E)=\sum_i\overline{\delta}(E\I{-}E_i^{\lambda})
\label{dens}\,,
\end{equation}
where $\overline{\delta}$ is a smoothed $\delta$ function (e.g. a Gaussian of width exceeding the typical spacing between levels), and the smoothed level flow 
\begin{equation}
\overline{j}^{\lambda}(E)=\overline{\rho}^{\lambda}(E)\ \overline{\phi}^{\lambda}(E)=\sum_i 
\frac{\partial E_i^{\lambda}}{\partial\lambda}\ \overline{\delta}(E\I{-}E_i^{\lambda})
\label{flow}\,,
\end{equation}
where $\overline{\phi}^{\lambda}(E)$ represents a \lq\lq velocity field\rq\rq\ of the spectrum.
It was shown in part~I \cite{ref:E1} that the smoothed level density and flow satisfy the ordinary continuity equation ensuring that levels do not disappear when $\lambda$ is varied. 
Substituting into Eq.\,(\ref{flow}) the Hellmann-Feynman formula,
\begin{equation}
\frac{\partial E_i^{\lambda}}{\partial\lambda}=\matr{\psi_i^{\lambda}}{\frac{\partial\hat{H}^{\lambda}}{\partial\lambda}}{\psi_i^{\lambda}}
\label{hell}\,,
\end{equation}
with $\ket{\psi_i^{\lambda}}$ standing for the eigenvector associated with level $E_i^{\lambda}$, we find that $\overline{j}^{\lambda}(E)\I{=}\overline{\partial_{\lambda}H}^{\lambda}(E)$, where the symbol on the right-hand side denotes the smoothed expectation value (strength function) of observable $\frac{\partial}{\partial\lambda}\hat{H}^{\lambda}$.

The above-introduced smoothed quantities play a crucial role in the definition and classification of ESQPTs.
The smoothed level density coincides with the semiclassical density of states obtained by integration of the classical Hamiltonian ${\cal H}^{\lambda}(\vecb{q},\vecb{p})$ over the phase space, namely 
\begin{equation}
\overline{\rho}^{\lambda}(E)\propto\frac{\partial}{\partial E}\iint\limits_{{\bf\Omega}} d^f\vecb{q}\,d^f\vecb{p}\ \Theta\left[E\I{-}{\cal H}^{\lambda}(\vecb{q},\vecb{p})\right]
\label{claden}\,,
\end{equation}
where $\Theta(x)$ represents the step function ($\Theta\I{=}0$ for $x\I{<}0$ and $\Theta\I{=}1$ for $x\I{\geq}0$) and ${\bf \Omega}$ is the domain of allowed $\{\vecb{q},\vecb{p}\}$ values.
A stationary point of the classical Hamiltonian yields a singularity of $\overline{\rho}^{\lambda}(E)$ as a function of energy, and this singularity propagates via Eqs.\,(\ref{flow}) and (\ref{hell}) also to energy dependencies of $\overline{j}^{\lambda}(E)$ and $\overline{\partial_{\lambda}H}^{\lambda}(E)$.

The classification of ESQPTs caused by non-degenerate stationary points (i.e., those with locally quadratic dependencies of ${\cal H}^{\lambda}$ on all $\vecb{q}$ and $\vecb{p}$ components) was given in Ref.\,\cite{ref:Stra16}.
The ESQPT non-analyticity associated  with such points appears in the $(f\I{-}1)$th derivative of the respective quantities and depends on the index $r$ of the stationary point (the number of negative eigenvalues of the Hessian matrix).
The $\lambda$-dependent energy ${\cal H}^{\lambda}=E^{\lambda}_{\rm c}$ of the stationary point forms a critical borderline (phase separatrix) in the plane $\lambda\times E$.
For systems with ${f=2}$, the non-degenerate stationary points cause singularities in the first derivative of the smoothed level density,
\begin{equation}
\frac{\partial\overline{\rho}^{\lambda}}{\partial E}\propto
\left\{\begin{array}{ll}
(-1)^{\frac{r}{2}}\ \Theta(E\I{-}E_{\rm c}^{\lambda})  & {\rm for}\ r\I{=}0,2,4,
\\
(-1)^{\frac{r+1}{2}}\ \ln|E\I{-}E_{\rm c}^{\lambda}| & {\rm for}\ r\I{=}1,3,
\end{array}\right.
\label{sinde}
\end{equation}
and in the first-derivative of the smoothed level slope:
\begin{equation}
\frac{\partial\overline{j}^{\lambda}}{\partial E}\I{\equiv}\frac{\partial\,\overline{\partial_{\lambda}H}^{\lambda}}{\partial E}\propto
\left\{\begin{array}{ll}
\Theta(E\I{-}E_{\rm c}^{\lambda})  & {\rm for}\ r\I{=}0,2,4,
\\
\ln|E\I{-}E_{\rm c}^{\lambda}| & {\rm for}\ r\I{=}1,3.
\end{array}\right.
\label{sinflo}
\end{equation}
Note that the sign prefactors in Eq.\,(\ref{sinflo}) are not determined and may differ from those in Eq.\,(\ref{sinde}).
Degenerate stationary points (flatter than quadratic) may cause even sharper singularities, which however are not generally classified \cite{ref:Stra16}.

Our intention is to perform an ESQPT analysis of general interacting boson systems introduced in Sec.\,\ref{sec:MBHa}.
We saw (cf.\,Table~\ref{gene}) that the classical-limit Hamiltonian ${\cal H}^{\lambda}$ of such a system with one- plus two-body interactions is a quartic function of coordinates and momenta $\{\vecb{q},\vecb{p}\}$ defined within a compact phase-space domain ${\bf\Omega}$ from Eq.\,(\ref{sphere}).
It turns out that to find stationary points of this setup is more complicated than in the cases studied in parts~I and II \cite{ref:E1,ref:E2} for two reasons:
First, the analysis cannot be reduced just to the coordinate space due to complicated kinetic terms of ${\cal H}^{\lambda}$.
Second, the compactness of the domain ${\bf\Omega}$ may create additional ESQPT singularities not connected to stationary points.
These issues are outlined in the following two subsections.

\subsubsection{Role of coordinate-dependent kinetic terms}
\label{sec:rol}

Stationary points of a common Hamiltonian ${{\cal H}^{\lambda}(\vecb{q},\vecb{p})={\cal K}^{\lambda}(\vecb{p})+{\cal V}^{\lambda}(\vecb{q})}$, with the kinetic term ${\cal K}^{\lambda}$ being a quadratic function of momenta and the potential ${\cal V}^{\lambda}$ an arbitrary function of coordinates, are always found at ${(\vecb{q},\vecb{p})=(\vecb{q}^{\lambda}_0,0)}$, where $\vecb{q}^{\lambda}_0$ are stationary points  of the potential.
This reduces the stationary-point analysis from the full $2f$-dimensional phase space just to the corresponding $f$-dimensional configuration space.
The study of ESQPTs benefits from this reduction, see parts~I and II \cite{ref:E1,ref:E2}.
However, Hamiltonians corresponding to generic interacting boson systems contain intermixed combinations of coordinates and momenta that typically yield coordinate-dependent kinetic terms.
This makes the analysis more complicated.

In general, the potential energy term ${\cal V}^{\lambda}(\vecb{q})$ is extracted from the total Hamiltonian ${\cal H}^{\lambda}(\vecb{q},\vecb{p})$ of any form by setting ${\dot{\vecb{q}}=0}$, so ${\frac{\partial}{\partial p_k}{\cal H}^{\lambda}=0}$ for all ${k=1,...,f}$.
This condition is satisfied for some momenta ${\vecb{p}=\vecb{p}_0^{\lambda}(\vecb{q})}$ and we can set ${{\cal V}^{\lambda}(\vecb{q})\equiv{\cal H}^{\lambda}(\vecb{q},\vecb{p}^{\lambda}_0(\vecb{q}))}$ and ${{\cal K}^{\lambda}(\vecb{q},\vecb{p})\equiv{\cal H}^{\lambda}(\vecb{q},\vecb{p})-{\cal V}^{\lambda}(\vecb{q})}$.
However, there may exist more than one solutions $\vecb{p}^{\lambda}_0(\vecb{q})$, which makes the above decomposition ambiguous.

The construction described in Sec.\,\ref{sec:MBHa} ensures that the classical limit of any bosonic Hamiltonian \eqref{Habo} contains only even powers of momentum $\vecb{p}$.
Therefore, the set of nonlinear equations following from the ${\dot{\vecb q}=0}$ condition has solutions of two types:
(i) the trivial solution ${\vecb{p}^{\lambda}_0=0}$, which is always present and independent of $\vecb{q}$, and (ii) non-trivial solutions, which come in sign conjugated pairs ${\vecb{p}_0^{\lambda}=\pm\vecb{p}_{0i}^{\lambda}(\vecb{q})}$ enumerated by index~$i$.
The latter solutions may exist only in a limited domain of $\vecb{q}$ or may be absent altogether, so their number cannot be set from general arguments.
The trivial solution is usually employed in the decomposition of the Hamiltonian to the kinetic and potential terms:
\begin{equation}
{\cal H}^{\lambda}(\vecb{q},\vecb{p})=
\underbrace{{\cal H}^{\lambda}(\vecb{q},\vecb{p})\I{-}{\cal H}^{\lambda}(\vecb{q},0)}_{{\cal K}^{\lambda}_0(\vecb{q},\vecb{p})}
+\underbrace{{\cal H}^{\lambda}(\vecb{q},0)}_{{\cal V}^{\lambda}_0(\vecb{q})}\,.
\label{kipo}
\end{equation}

Specifying the solution $\vecb{p}^{\lambda}_0(\vecb{q})$ of $\dot{\vecb{q}}\I{=}0$, one needs to complete the determination of stationary points by solving the ${\dot{\vecb p}=0}$ condition, so ${\frac{\partial}{\partial q_k}{\cal H}^{\lambda}=0}$ for all $k=1,...,f$.
This is certainly achieved by the usual procedure of finding $\vecb{q}^{\lambda}_0$ such that 
\begin{equation}
\left.\frac{\partial}{\partial q_k}{\cal V}^{\lambda}_0(\vecb{q})\right|_{\vecb{q}=\vecb{q}^{\lambda}_0}\I{=}0
\,.
\label{statri}
\end{equation}
Indeed, the points ${(\vecb{q},\vecb{p})=(\vecb{q}^{\lambda}_0,0)}$ are stationary points of the whole Hamiltonian as we obviously have ${\frac{\partial}{\partial q_k}{\cal K}^{\lambda}_0=0}$ at $\vecb{p}\I{=}0$ for any $\vecb{q}$.
However, within the domains of existence of the non-trivial solutions $\pm\vecb{p}_{0i}^{\lambda}(\vecb{q})$ of ${\dot{\vecb q}=0}$, there may be some additional solutions of the ${\dot{\vecb p}=0}$ problem.
These read as ${(\vecb{q},\vecb{p})=(\vecb{q}_{0i}^{\lambda},\pm\vecb{p}_{0i}^{\lambda}(\vecb{q}_{0i}^{\lambda}))}$ and have to satisfy
\begin{equation}
\left.\frac{\partial {\cal K}^{\lambda}_0(\vecb{q},\vecb{p})}{\partial q_k}\right|_{\begin{smallmatrix}\vecb{q}=\vecb{q}_{0i}^{\lambda}\qquad\\\vecb{p}=\pm\vecb{p}_{0i}^{\lambda}(\vecb{q}_{0i}^{\lambda})\end{smallmatrix}}
\!\!\!\!\!\!\!+
\left.\frac{\partial {\cal V}^{\lambda}_0(\vecb{q})}{\partial q_k}\right|_{\vecb{q}=\vecb{q}_{0i}^{\lambda}}
\!\!\!=0
\label{stanotri}
\end{equation}
for all ${k=1,...,f}$.
This condition follows directly from the full Hamiltonian decomposition in Eq.\,(\ref{kipo}).

We can therefore conclude that Hamiltonians with coordinate-dependent kinetic terms may generate stationary points beyond the analysis of the potential energy term alone.
In the study of ESQPTs in such systems, both Eqs.\,(\ref{statri}) and (\ref{stanotri}) must be taken into account.

\subsubsection{Effects of phase-space boundary}
\label{sec:bou}

The boundedness of the phase space implies the existence of an upper limit $E^{\lambda}_{\rm max}$ of the energy spectrum, which for $N\to\infty$ coincides with the maximal value ${\cal H}^{\lambda}_{\rm max}$ of the classical Hamiltonian within the domain ${\bf\Omega}$.
For ${E>{\cal H}^{\lambda}_{\rm max}}$, formula~(\ref{claden}) ensures ${\overline{\rho}^{\lambda}(E)=0}$. 
It is therefore legitimate to expect that the level density $\overline{\rho}^{\lambda}(E)$ exhibits the following broad behavior: it grows from zero at the minimal (ground-state) energy $E^{\lambda}_{\rm min}$, reaches a maximum somewhere in the middle of the spectrum, and then decreases back to zero above $E^{\lambda}_{\rm max}$. 
Note that neither the increasing, nor the decreasing part of $\overline{\rho}^{\lambda}(E)$ must be monotonic.
We know (see Secs.\,\ref{sec:cla} and \ref{sec:rol}) that if the domain ${\bf\Omega}$ contains stationary points of ${\cal H}^{\lambda}$, they generate the respective types of ESQPTs.
Here we are going to show that the phase space boundary itself, i.e., the hypersurface ${\bf\partial\Omega}$, may create some extra singularities.

To analyze this effect, we introduce in the $2f$-dimensional phase space the set of hyperspherical coordinates ${R\in[0,\infty)}$ (radius) and ${\{\phi_l\}_{l=1}^{2f-1}\equiv\vecb{\phi}}$ (angles).
The energy landscape derived from the classical Hamiltonian ${\cal H}^{\lambda}(\vecb{q},\vecb{p})$ for ${R\in[0,R_0]}$ (i.e., inside ${\bf\Omega}$) then reads:
\begin{eqnarray}
E^{\lambda}(R,\vecb{\phi})=a^{\lambda}(R_0^2\I{-}R^2)\I{+}b^{\lambda}(\vecb{\phi})R(R_0^2\I{-}R^2)^{\frac{1}{2}}\I{+}c^{\lambda}(\vecb{\phi})R^2
\nonumber\\
\qquad + d^{\lambda}(\vecb{\phi}) R^4\I{+}e^{\lambda}(\vecb{\phi}) R^3(R_0^2\I{-}R^2)^{\frac{1}{2}}\I{+}f^{\lambda}(\vecb{\phi}) R^2(R_0^2\I{-}R^2)
\nonumber\\
+ g^{\lambda}(\vecb{\phi}) R(R_0^2\I{-}R^2)^{\frac{3}{2}}+h^{\lambda}(R_0^2\I{-}R^2)^2.
\,\,\ \qquad\qquad\qquad
\label{hyp}
\end{eqnarray} 
Here, $a^{\lambda}$, $b^{\lambda}$, ..., $h^{\lambda}$ are some coefficients (constants or functions of angles $\vecb{\phi}$), which are uniquely determined from the set of constants $\varepsilon^{\lambda}_{kk'}$ and $v^{\lambda}_{kk'll'}$ in the quantum Hamiltonian (\ref{Habo}).
Note that one- and two-body parts of the Hamiltonian yield terms $\sim R^2$ and $\sim R^4$, respectively, and that $(R_0^2-R^2)^{\frac{1}{2}}$ replaces both creation and annihilation operators of the $b_0$ boson.
Values $R\in(R_0,\infty)$ (outside ${\bf\Omega}$) are forbidden and one can visualize this by embedding the system into a sharp energy well yielding the energy $E^{\lambda}(R,\vecb{\phi})$ equal to (\ref{hyp}) for ${R\in[0,R_0]}$ and to infinity for ${R\in(R_0,\infty)}$.

On the hypersurface ${\bf\partial\Omega}$, the energy (\ref{hyp}) takes values 
\begin{equation}
E^{\lambda}_{\bf\partial\Omega}(\vecb{\phi})=c^{\lambda}(\vecb{\phi})\,R_0^2+d^{\lambda}(\vecb{\phi})\,R_0^4,
\label{E0}
\end{equation}
which depend only on angles $\vecb{\phi}$.
It turns out that the boundary generates a spectral singularity if there exists a point $\vecb{\phi}^{\lambda}_0$ such that ${\frac{\partial}{\partial\phi_l}E^{\lambda}_{\bf\partial\Omega}(\vecb{\phi})|_{\vecb{\phi}=\vecb{\phi}^{\lambda}_0}=0}$ for all ${l=1,...,2f-1}$.
At this point, whose energy is ${E^{\lambda}_{\bf\partial\Omega}(\vecb{\phi}^{\lambda}_0)\equiv E^{\lambda}_{\rm c}}$, the phase-space integration in Eq.\,(\ref{claden}) yields a non-analytic contribution to the level density $\overline{\rho}^{\lambda}(E)$.
We stress that the point $(R,\vecb{\phi})\I{=}(R_0,\vecb{\phi}^{\lambda}_0)$ does not have to be a stationary point of the whole Hamiltonian as we set no constraint on $\frac{\partial}{\partial R}E^{\lambda}(R,\vecb{\phi})$.

For the sake of concreteness, let us consider a local extreme (minimum or maximum) of the surface energy function $E^{\lambda}_{\bf\partial\Omega}(\vecb{\phi})$.
The extreme is assumed to be separable, i.e., expressed through ${E^{\lambda}_{\bf\partial\Omega}\propto\sum_{l=1}^{2f-1}(\delta r_l)^{K_l}}$, where $\delta r_l$ denote some local orthogonal coordinates on ${\bf\partial\Omega}$ in a vicinity of $\vecb{\phi}^{\lambda}_0$.
The powers $K_l$ specify energy dependencies in the respective directions: ${K_l=2}$ indicates a quadratic extreme, ${K_l=4}$ quartic etc.
${K_l\to\infty}$ leads to a totally flat (locally constant) dependence.
We introduce an average inverse power $K$ given by: ${K^{-1}=\sum_{l=1}^{2f-1}K_l^{-1}/(2f-1)}$.
It can be shown (the analysis is not presented here) that as the energy $E$ crosses the energy $E^{\lambda}_{\rm c}$ of the extreme, the smoothed level density close to $E^{\lambda}_{\rm c}$ varies according to
\begin{eqnarray}
\overline{\rho}^{\lambda}(E)=\varrho^{\lambda}(E)\left[1-\sigma\,\Theta\left[\pm(E\I{-}E_{\rm c}^{\lambda})\right]\abs{E-E^{\lambda}_{\rm c}}^I\,\right],
\nonumber\\
I=\frac{2f-1}{K}+\frac{1}{M}-1,\quad
\label{expo}
\end{eqnarray}
where $\varrho^{\lambda}(E)$ is an unspecified analytic function and ${\sigma={\rm sgn}\frac{\partial}{\partial R}E^{\lambda}|_{(R,\vecb{\phi})=(R_0,\vecb{\phi}_0^{\lambda})}}$. 
The sign $+$ or $-$, respectively, corresponds to a local minimum or maximum of the surface energy function and $M\I{\in}\{\frac{1}{2},1,\frac{3}{2},2\}$ stands for the lowest power in the dependence of $E^{\lambda}$ on ${(R-R_0)}$ at ${\vecb{\phi}=\vecb{\phi}^{\lambda}_0}$.
This indicates that $\overline{\rho}^{\lambda}(E)$ exhibits at $E^{\lambda}_{\rm c}$ a non-analyticity connected with  its $\lceil I\rceil$th derivative, where $\lceil x\rceil$ denotes the ceiling function.
The derivative is discontinuous if $I$ is integer, or divergent otherwise.
Note that for ${K=2}$ and ${M=\frac{1}{2}}$ (assumingly the most generic case as it does not require an accidental vanishing of the lowest order expansion coefficients in any of the energy dependences) we expect a divergent derivative $\frac{\partial^{f+1}}{\partial E^{f+1}}\overline{\rho}^{\lambda}(E)|_{E=E^{\lambda}_{\rm c}}$. 
On the other hand, the most \uvo{conspirative} case ${K=\infty}$ and ${M\geq 1}$ leads to discontinuous or divergent $\overline{\rho}^{\lambda}(E)|_{E=E^{\lambda}_{\rm c}}$, independently of $f$.

The analysis of other types of stationary points of the boundary function $E^{\lambda}_{\bf\partial\Omega}(\vecb{\phi})$ needs to be performed case by case.
Note that the energy $E^{\lambda}_{\bf\partial\Omega}(\vecb{\phi}_0)$ may coincide with that of a regular stationary point inside ${\bf\Omega}$, which enhances the corresponding  singularity in the spectrum.

\subsubsection{Finite-size effects due to partial separability}
\label{sec:sep}

In part~II \cite{ref:E2} we have shown that finite-size effects in the oscillatory component $\widetilde{\rho}^{\lambda}(E)$ of the level density can be particularly strong in ${f=2}$ systems that exhibit partial (effective) decoupling of both degrees of freedom.
If one degree of freedom (mode of motion) in such a decoupled system is much faster than the other, the quantum energy spectrum consists of separated bands of states corresponding to distinguished excitations of both modes. 
Excitations of the fast mode form a sequence of levels with large spacings, each of them being a base for excitations of the slow mode that give rise to repeated bands of levels with small spacings.
If, in such a situation, one of the modes undergoes an ESQPT, we observe a peculiar pattern of finite-size precursors that carry strong features of the underlying ${f=1}$ dynamics.
A qualitative analysis of these phenomena has been presented in terms of an effective Hamiltonian describing one-dimensional dynamics based on a boosted momentum of the other degree of freedom, see formula (28) in part~II \cite{ref:E2}.

In a general interacting boson system, these considerations can be put on a more rigorous ground with the aid of generalized projective coherent states, which are outlined in Appendix~\ref{ApB}.
Suppose that the fast mode of motions of the system, that is at least partially separated from the other(s), is associated with the boson creation operator $\hat{B}^{\dag}(\tilde{\alpha}_{\perp})$.
This boson plays the role of a single-mode excitation quantum and must be perpendicular to the ground-state condensate boson, represented by operator $\hat{B}^{\dag}(\tilde{\alpha})$ from Eq.\,\eqref{proco}.
Therefore, in absence of other types of excitations, the states of the system with the total number of bosons $N$ have the form $\propto\hat{B}^{\dag}(\tilde{\alpha}_{\perp})^{N_{\perp}}\hat{B}^{\dag}(\tilde{\alpha})^{N\I{-}N_{\perp}}\ket{0}$, where $N_{\perp}$ is the number of excitation quanta.

We know that parameters $\tilde{\alpha}$ from the expansion of $\hat{B}^{\dag}(\tilde{\alpha})$ are associated with coordinates and momenta of the system.
The classical counterpart associated with a quantum Hamiltonian $\hat{H}^{\lambda}$ is expressed solely in terms of the ground-state boson $\hat{B}^{\dag}(\tilde{\alpha})$, namely as ${{\cal H}^{\lambda}(\tilde{\alpha},N)=\matr{0}{\hat{B}(\tilde{\alpha})^N\hat{H}^{\lambda}\hat{B}^{\dag}(\tilde{\alpha})^N}{0}/(N!N) }$, see Eq.\,\eqref{Haproco}. 
An excited energy surface, which underlies motions of any character on top of the state with $N_{\perp}$ excitation quanta of the above type, can be written as:
\begin{eqnarray}
{\cal H}^{\lambda}_{\rm ex}(\tilde{\alpha},N\I{-}N_{\perp},N_{\perp})=
\qquad\qquad\qquad\qquad\qquad
\nonumber\\
\frac{\matr{\tilde{\alpha}_{\perp},N_{\perp}}{\hat{B}(\tilde{\alpha})^{N-N_{\perp}}\hat{H}^{\lambda}\hat{B}^{\dag}(\tilde{\alpha})^{N-N_{\perp}}}{\tilde{\alpha}_{\perp},N_{\perp}}}{(N\I{-}N_{\perp})!\,N},
\label{Haprocoex}
\end{eqnarray}
where ${\ket{\tilde{\alpha}_{\perp},N_{\perp}}=\frac{1}{\sqrt{N_{\perp}!}}\hat{B}^{\dag}(\tilde{\alpha}_{\perp})^{N_{\perp}}\ket{0}}$.
Since coefficients $\tilde{\alpha}_{\perp}$ depend via the orthogonality constraint on $\tilde{\alpha}$, the above expression is written as a function of only the latter variables, i.e., standard coordinates and momenta.
Formula \eqref{Haprocoex} is a special case of Eq.\,\eqref{expos}.
It will be applied below in the analysis of oscillatory structures in the spectrum of the IBM.

\section{The IBM as a case study}
\label{sec:IBM}

\subsection{First-order QPT Hamiltonian }

In what follows, we explore the general ESQPT signatures in the the interacting boson model of collective nuclear dynamics \cite{ref:IBMbook}.
The model, describing rotations and quadrupole vibrations of nuclei, is formulated in terms of bosons $s$ and $d$ with angular momenta 0 and 2, respectively, and has ${f=5}$ degrees of freedom.
Considering bilinear combinations of boson creation and annihilation operators, one builds the U(6) dynamical group which branches into three dynamical-symmetry chains starting with subgroups U(5), O(6) and SU(3).
All Hamiltonians are rotationally invariant under the group O(3), which ensures conservation of the total angular momentum quantum number $L$.
The conserved total number of bosons $N$, resulting from selection of a single irrep of U(6), can undergo the limiting process ${N\to\infty}$, which leads to semiclassical description and sharpening of quantum critical effects.
 
The IBM was instrumental in setting the foundations for both QPT and ESQPT phenomena \cite{ref:Diep80,ref:Feng81,ref:Cejn09}.
The original papers on ESQPTs \cite{ref:Cejn06,ref:Hein06,ref:Mace06} considered the transition between spherical and axially-unstable quadrupole shapes of the ground state [U(5) and O(6) dynamical symmetries].
In this transition, the model is fully integrable all the way and exhibits a second-order ground-state QPT.
The seniority quantum number $\tau$ is conserved and the constraint ${\tau=L=0}$ reduces the number of effective degrees of freedom to ${f=1}$.
This makes the ESQPT signatures very well visible---they affect in the same way all individual energy levels satisfying the above constraint and are apparent already in the zeroth derivative of the smoothed level density and slope.

Here we focus on the transition between spherical and axially-deformed ground-state shapes, which requires a more complex treatment.
The irrotational constraint $L=0$ yields the effective number of degrees of freedom ${f=2}$ (no additional conserved quantity except $N$ exists), which implies that the ESQPT spectral singularities appear in the first derivative of the smoothed level density and slope.
Studying the ESQPTs accompanying the first-order ground-state QPT, we parallel the ESQPT analysis done in parts~I and II \cite{ref:E1,ref:E2} and extend it into the setting of a realistic interacting many-body system. 

The structural evolution between spherical and axially-deformed equilibrium configurations of the IBM was studied from the ESQPT perspective in Refs.\,\cite{ref:Zhan16,ref:Zhan17}.
The Hamiltonian used there was of the well known form
\begin{equation}
\hat{H}^{(\eta,\chi)}=(1\I{-}\eta)\,\hat{n}_d-\frac{\eta}{4N}(\hat{Q}^{\chi}\cdot\hat{Q}^{\chi})
\label{Haha}
\end{equation}
depending on parameters ${\eta\in[0,1]}$ and ${\chi\in[-\frac{\sqrt{7}}{2},+\frac{\sqrt{7}}{2}]}$.
Here, ${\hat{n}_d=(\hat{d}^{\dag}\cdot\hat{\tilde{d}})}$ is the $d$-boson number operator and ${\hat{Q}^{\chi}_{\mu}=[\hat{d}^{\dag}\hat{s}+\hat{s}^{\dag}\hat{\tilde{d}}]^{(2)}_{\mu}+\chi[\hat{d}^{\dag}\hat{\tilde{d}}]^{(2)}_{\mu}}$ the quadrupole operator.
For ${\chi=-\frac{\sqrt{7}}{2}}$, ${\chi=+\frac{\sqrt{7}}{2}}$ or ${\chi=0}$, the Hamiltonian depending on $\eta$ describes the U(5)--SU(3), U(5)--$\overline{{\rm SU(3)}}$ or U(5)--O(6) transition, respectively, while for ${\eta=1}$, the Hamiltonian depending on $\chi$ describes the $\overline{{\rm SU(3)}}$--O(6)--SU(3) transition. 
Note that $\overline{{\rm SU(3)}}$ differs from SU(3) by the type of equilibrium deformation (oblate instead of prolate).
We use the symbol ${[\hat{A}^{(\ell_1)}\hat{B}^{(\ell_2)}]^{(\ell)}_{m}}$ for the coupling of spherical tensors with angular momenta $\ell_1$ and $\ell_2$ to one with angular momentum $\ell$ and projection $m$, while the scalar product is written as ${(\hat{A}^{(\ell)}\cdot\hat{B}^{(\ell)})=(-1)^{\ell}\sqrt{2\ell+1}[\hat{A}^{(\ell)}\hat{B}^{(\ell)}]^{(0)}_0}$. 
The tilde above an annihilation operator means ${\hat{\tilde{x}}^{(\ell)}_{m}=(-1)^{\ell+m}\hat{x}^{(\ell)}_{-m}}$, ensuring proper transformation properties under rotations.

In this paper, we employ a more sophisticated Hamiltonian adopted from the formalism resolving the intrinsic and collective motions within the IBM, see Refs.\,\cite{ref:Kirs85,ref:Levi87}.
This Hamiltonian enables us to generate a ground-state shape-phase transition with an adjustable barrier between the spherical and deformed configurations \cite{ref:Levi06,ref:Mace11,ref:Levi12,ref:Mace14}, which is in contrast to the \lq\lq consistent-$Q$\rq\rq\ Hamiltonian \eqref{Haha} yielding a very small barrier.
In particular, we use two Hamiltonian forms,
\begin{eqnarray}
\hat{H}_1(\beta_0',\zeta)&=&\frac{2(1\I{-}\zeta^2)}{N}\ \hat{n}_d(\hat{n}_d\I{-}1)+\frac{1}{N}\,\hat{D}^{\dag}(\beta_0',\zeta)\I{\cdot}\hat{\tilde{D}}(\beta_0',\zeta)
\nonumber\\
\label{H1n}
\\
\hat{H}_2(\beta_0',\xi)&=&\hat{H}_1(\beta_0',\zeta\I{=}1)+\frac{\xi}{N}\,\hat{S}^{\dag}(\beta_0')\hat{S}(\beta_0')
\,,
\label{H2n}
\end{eqnarray}
where $\zeta\in[0,1]$ and $\xi\in[0,1]$ are two control parameters and $\hat{S}^{\dag}(\beta_0')$ and $\hat{D}^{\dag}(\beta_0',\zeta)$ two specific boson pair operators with ${\ell=0}$ and 2:
\begin{eqnarray}
\hat{S}^{\dag}(\beta_0')&=&\hat{d}^{\dag}\I{\cdot}\hat{d}^{\dag}-\beta_0'^\,\hat{s}^{\dag}\hat{s}^{\dag}
\label{S}
\,,\\
\hat{D}_{\mu}^{\dag}(\beta_0',\zeta)&=&\sqrt{2}\beta_0'\bigl[\hat{s}^{\dag}\hat{d}^{\dag}\bigr]^{(2)}_{\mu}+\sqrt{7}\zeta\bigl[\hat{d}^{\dag}\hat{d}^{\dag}\bigr]^{(2)}_{\mu}
\label{D}
\,.
\end{eqnarray}
$\beta_0'$ is an adjustable parameter that modifies the shape of the potential energy functions associated with Hamiltonians \eqref{H1n} and \eqref{H2n}.
It affects the size of the equilibrium deformation, rigidity of spherical and deformed potential wells, and the height of the barrier between these configurations in the phase-transitional domain.
Note that for the purposes of the present paper we modify the original notation of Refs.\,\cite{ref:Mace11,ref:Levi12,ref:Mace14}, where the control parameter of Eq.\,\eqref{H1n} is denoted as ${\rho\equiv\zeta/\beta_0'}$ while operators ${\hat{P}_0^{\dag}(\beta_0')\equiv\hat{S}^{\dag}}(\beta_0')$, ${\hat{R}_{2\mu}^{\dag}(\rho)\equiv\hat{D}_{\mu}^{\dag}(\beta_0',\zeta=\beta_0'\rho)/\beta_0'}$ and ${\hat{P}^{\dag}_{2\mu}(\beta_0')\equiv\hat{D}_{\mu}^{\dag}(\beta_0',\zeta=1)}$ are employed instead of those in Eqs.\,\eqref{S} and \eqref{D}.

Hamiltonian ${\hat{H}_1(\beta_0',\zeta=0)}$ represents the U(5) limit of the model, while ${\hat{H}_2(\beta_0'=\sqrt{2},\xi=1)}$ corresponds to the SU(3) limit.
The first-order critical-point Hamiltonian between these two dynamical symmetries reads ${\hat{H}_{\rm c}(\beta_0')\equiv\hat{H}_1(\beta_0',\zeta=1)=\hat{H}_2(\beta_0',\xi=0)}$.
Hence the transition from the U(5) limit to the critical point is realized by Hamiltonian (\ref{H1n}) with $\zeta$ varying from 0 to~1, and the subsequent transition from the critical point to the SU(3) limit is driven by Hamiltonian (\ref{H2n}) with $\xi$ changing from 0 to~1 and $\beta_0'$ converging to $\sqrt{2}$.
In agreement with the notation used in Sec.\,\ref{sec:MB}, we introduce a single control parameter $\lambda\in[0,\infty)$ and define the Hamiltonian as follows:
\begin{equation}
\hat{H}^{\lambda}(\beta_0')=\left\{\begin{array}{ll}
\hat{H}_1(\beta_0',\zeta\I{=}\lambda) & {\rm for\ }\lambda\in[0,1],\\
\hat{H}_2(\beta_0',\xi\I{=}\lambda\I{-}1) & {\rm for\ }\lambda\in[1,\infty).
\end{array}\right.
\label{Hamib}
\end{equation}
Since $\beta_0'$ is not necessarily restricted to $\sqrt{2}$ and $\lambda$ may exceed the value 2 (corresponding to ${\xi=1}$), we do not insist on the SU(3) dynamical symmetry as an endpoint of the transition.
We stress that although both parts of this formula coincide at the critical point ${\lambda=1\equiv\lambda_{\rm c}}$, implying that all eigenstates are the same there, the slopes $\frac{\partial}{\partial\lambda}E_i^{\lambda}$ of excited levels differ as one approaches to $\lambda_{\rm c}$ from the right and from the left.
The energy of the ground state is set to ${E_0^{\lambda}=0}$ independently of $\lambda$.

The Hamiltonian described above was studied in detail from the viewpoint of classical and quantum chaos and with respect to partial and quasi dynamical symmetries \cite{ref:Mace11,ref:Levi12,ref:Mace14}.
Its QPT and ESQPT analysis relies upon the construction of the classical limit.
With the aid of the Glauber coherent-state formalism in the form presented in Ref.\,\cite{ref:Hatc82} (cf. Sec.\,\ref{sec:MBHa} above) the classical counterparts of Hamiltonians (\ref{H1n}) and (\ref{H2n}) constrained to $L=0$ can be cast in the following way,
\begin{widetext}
\begin{eqnarray}
\label{eq:H1cl}
\mathcal{H}_1 (\beta_0',\zeta)&\I{=}&
{\cal H}_{d,0}^2 + \bz^2(1\I{-}{\cal H}_{d,0}){\cal H}_{d,0}
+ \zeta^2  p_\gamma^2
+ \zeta\bz\sqrt{\tfrac{1}{2}(1\I{-}{\cal H}_{d,0})}
\left [\,(\beta^{-1} p_\gamma^2\I{-}\beta p^2_{\beta}\I{-}\beta^3) \cos{3\gamma}
+ 2p_{\beta} p_\gamma \sin{3\gamma}\,\right],
\\
\label{eq:H2cl}
\mathcal{H}_2(\beta_0',\xi)&\I{=}& 
\mathcal{H}_1 (\beta_0',\zeta=1)
+\tfrac{1}{2}\ \xi \left [\,\beta^2 p_\beta^2 + \tfrac{1}{4}(\beta^2\I{-} T)^2
- \bz^2(1 - {\cal H}_{d,0})(\beta^2\I{-}T)
+ \bz^4(1\I{-}{\cal H}_{d,0})^2\,\right],
\end{eqnarray}
\end{widetext}
where $T = p_{\beta}^2+\beta^{-2}p_\gamma^2$ denotes the standard kinetic term and ${\cal H}_{d,0}=\tfrac{1}{2}(T+\beta^2)$ stands for the classical limit of $\hat{n}_d$ restricted to $L=0$.

Polar coordinates ${\beta\in[0,\sqrt{2}]}$ and ${\gamma\in[0,2\pi)}$ are related to the Bohr deformation variables, and can be represented by equivalent Cartesian coordinates ${x=\beta\cos\gamma}$ and ${y=\beta\sin\gamma}$.
Together with the canonically conjugate momenta ${p_\beta\in[0,\sqrt{2}]}$ and ${p_\gamma\in[0,1]}$, or ${p_x=p_{\beta}\cos\gamma-\frac{p_{\gamma}}{\beta}\sin\gamma\in[0,\sqrt{2}]}$ and ${p_y=p_{\beta}\sin\gamma+\frac{p_{\gamma}}{\beta}\cos\gamma\in[0,\sqrt{2}]}$, they span a compact classical phase space domain ${\bf\Omega}$, in agreement with Eq.\,\eqref{sphere}.
The Hamiltonian exhibits the well known discrete symmetry under the shift ${\gamma\to\gamma+\frac{2\pi}{3}}$. 
Note that the present bounded form of the deformation parameter $\beta$ can be transformed to a boundless form ${\beta'=\beta/\sqrt{2-\beta^2}\in[0,\infty)}$ \cite{ref:Klei81}, which is more common in some literature.
The value ${\beta_0'=\sqrt{2}}$ of the shape parameter is associated with the equilibrium deformation ${\beta_0=2/\sqrt{3}}$ in the SU(3) limit.

\subsection{ESQPT analysis}
\label{sec:Res}

\subsubsection{Smooth level density}
\label{sec:Ressmo}

\begin{figure}[ht!]
\includegraphics[width=\linewidth]{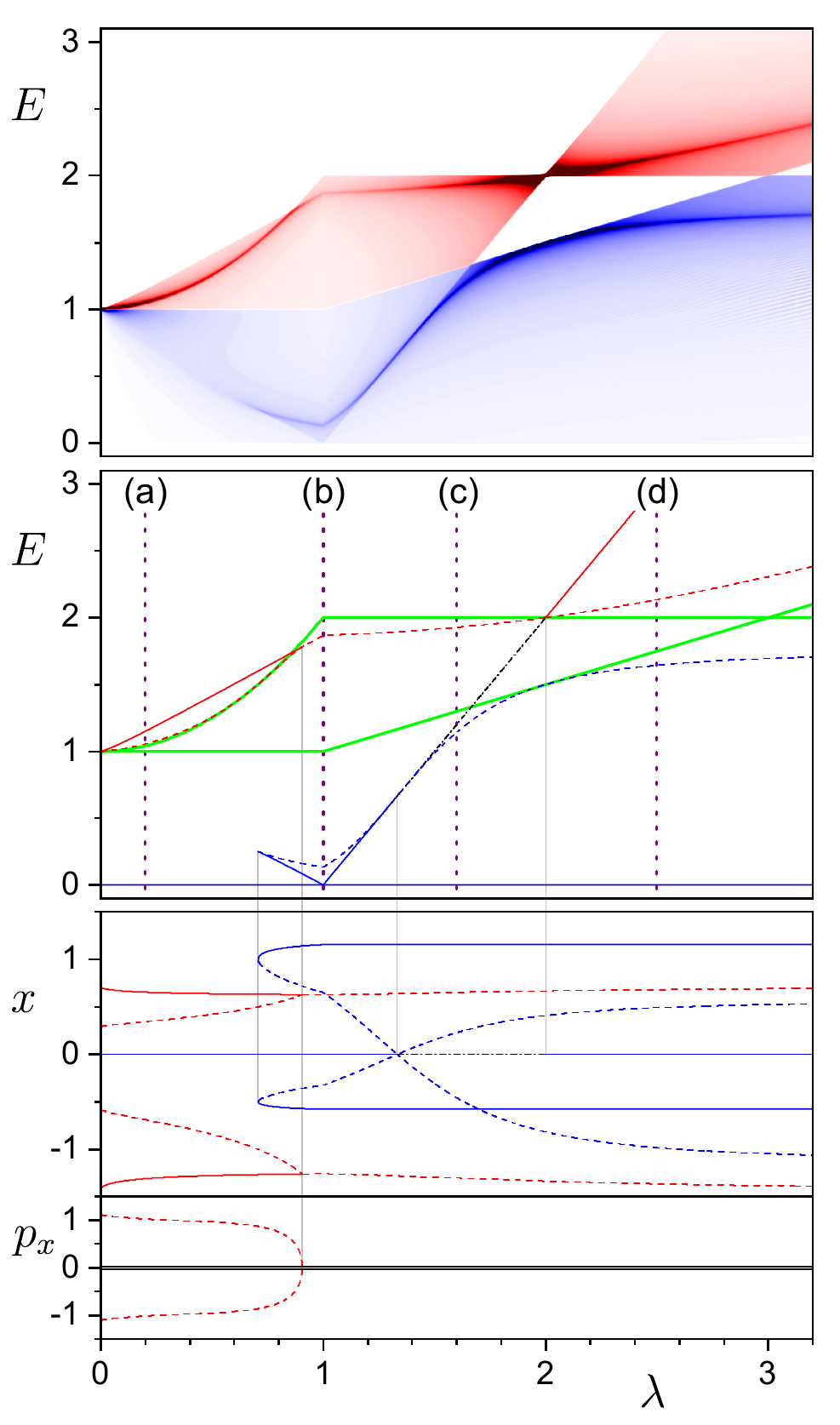}
\caption{Excited-state quantum phase diagram and classical stationary points of the IBM Hamiltonian \eqref{Hamib} with ${\beta_0'=\sqrt{2}}$.
The top panel shows density plots of the first derivative $\frac{\partial}{\partial E}\overline{\rho}^{\lambda}(E)$ of the smooth level density as a function of control parameter $\lambda$ and energy $E$ (the darker shades represent larger absolute values).
The blue and red areas indicate positive and negative derivatives, respectively, white areas correspond to zero derivatives. 
The middle panel displays the ESQPT critical borderlines, as described in the text [see items (i)--(vi)].
Vertical lines (a)--(d) demarcate the cuts shown in Fig.\,\ref{f:Hus1}.
The bottom panel depicts the coordinate ${x=\beta\cos\gamma}$ and momentum $p_x$ demarcating $\lambda$-dependendent positions of stationary points in the classical phase space.
The line types correspond to those indicating the respective ESQPT borderlines in the middle panel and in Fig.\,\ref{f:Hus1}. 
The thick black line at ${p_x=0}$ in the bottom panel contains all remaining lines.
All axes are dimensionless.}
\label{f:Dia1}
\end{figure}   

\begin{figure}[ht!]
\includegraphics[width=\linewidth]{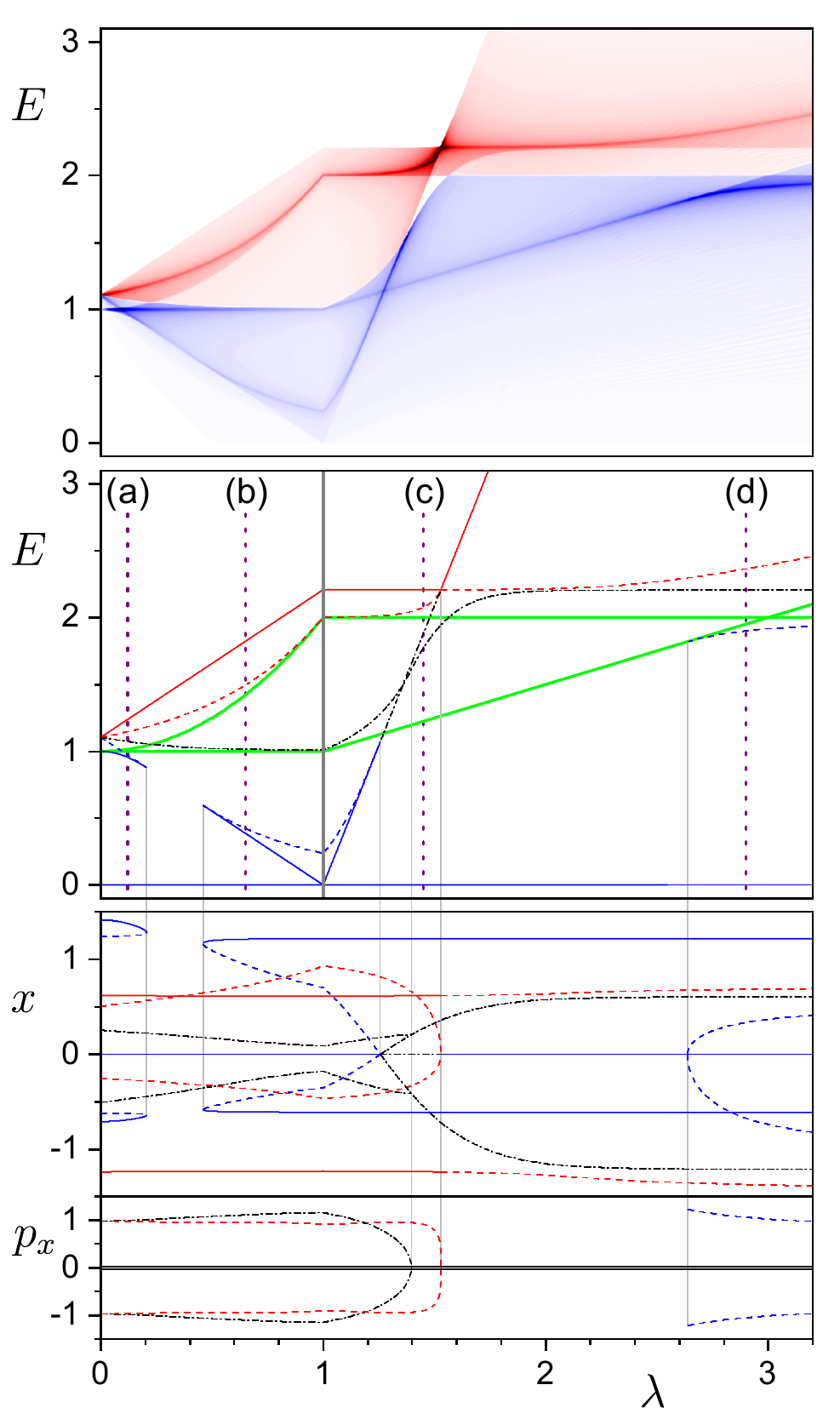}
\caption{The same as in Fig.\,\ref{f:Dia1}, but for ${\beta_0'=1.7}$, which yields a richer phase structure.
Vertical lines (a)--(d) demarcate the cuts of the level density described in Fig.\,\ref{f:Hus2}.}
\label{f:Dia2}
\end{figure}   

\begin{figure}[t!]
\includegraphics[width=\linewidth]{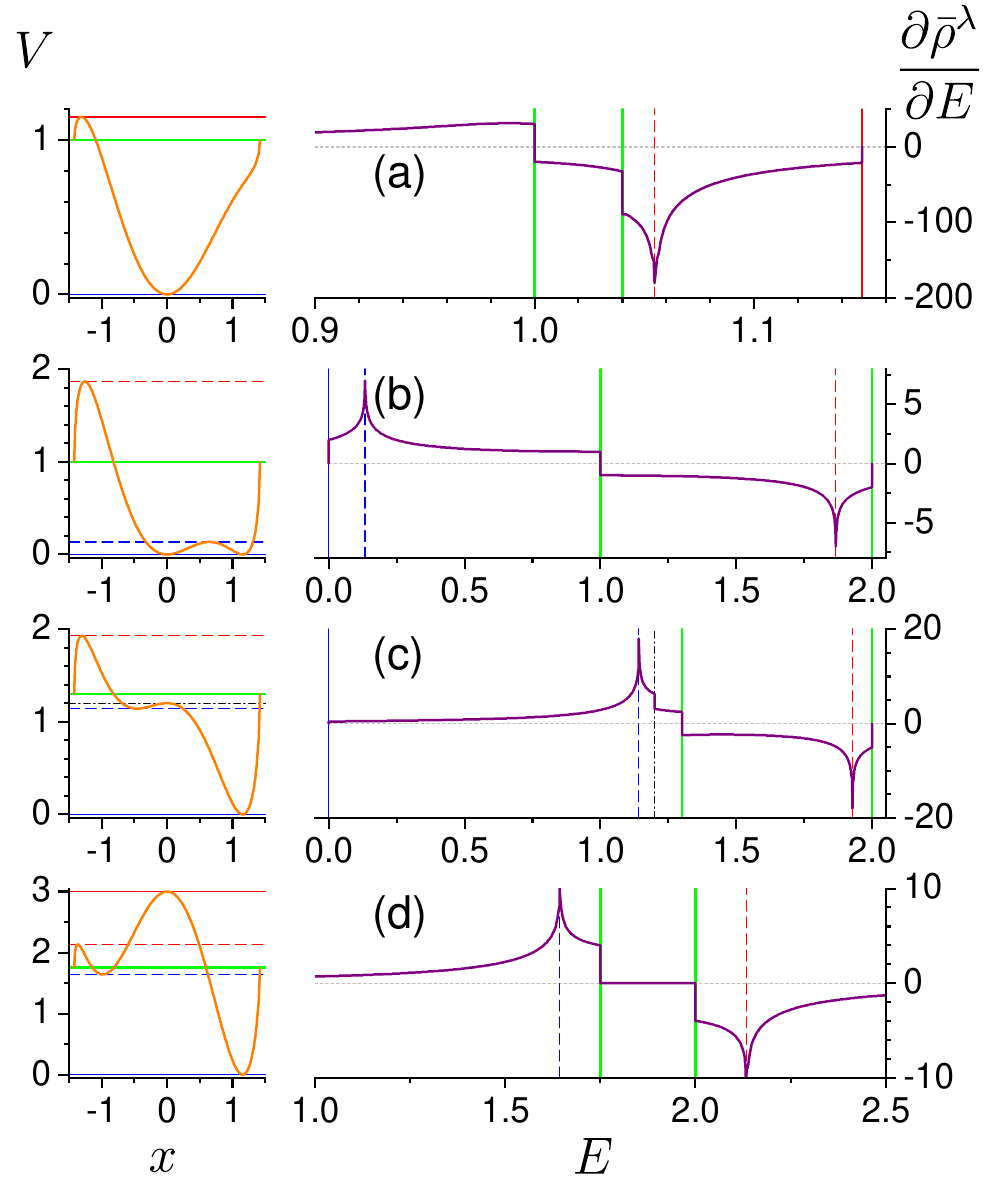}
\caption{
The potential energy function ${{\cal V}^{\lambda}(x,y=0)}$ (left column) and the level density derivative $\frac{\partial}{\partial E}\overline{\rho}^{\lambda}(E)$ (right column) for four values of parameter $\lambda$ in Hamiltonian \eqref{Hamib} with ${\beta_0'=\sqrt{2}}$. 
Rows (a)--(d) correspond to the respective cuts at ${\lambda=0.2, 1.0, 1.6}$ and 2.5  in Fig.\,\ref{f:Dia1}.
Recall: ${(\lambda^*,\lambda_{\rm c},\lambda^{**})=(0.707,1,1.333)}$.
Energies of ESQPTs are indicated by the same types of horizontal and vertical lines in the left and right columns, respectively, also in correspondence to Fig.\,\ref{f:Dia1}.
All axes are dimensionless.}
\label{f:Hus1}
\end{figure}   

\begin{figure}[t!]
\includegraphics[width=\linewidth]{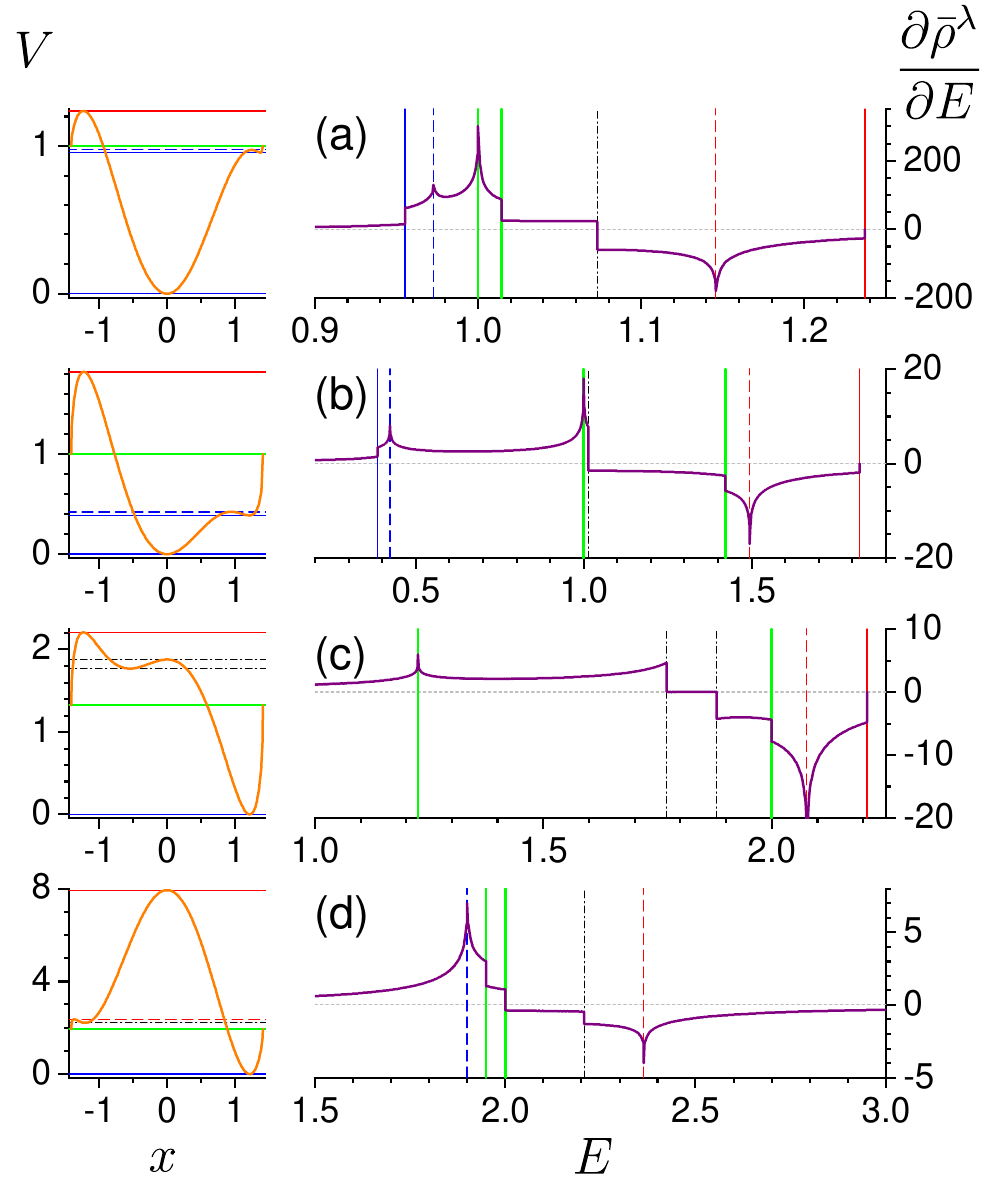}
\caption{The same as in Fig.\,\ref{f:Hus1}, but for ${\beta_0'=1.7}$.
Rows (a)--(d) correspond to the respective cuts at ${\lambda = 0.12, 0.65, 1.45}$ and 2.90 in Fig.\,\ref{f:Dia2}.
Recall: ${(\lambda^*,\lambda_{\rm c},\lambda^{**})=(0.460,1,1.257)}$.
}
\label{f:Hus2}
\end{figure}   

Results of our numerical analysis of the smooth component $\overline{\rho}^{\lambda}(E)$ of the level density associated with the IBM Hamiltonian \eqref{Hamib} are shown in Figs.\,\ref{f:Dia1}--\ref{f:Hus2}.
We use two values of the Hamiltonian shape parameter $\beta_0'$, namely ${\beta_0'=\sqrt{2}}$ in Figs.\,\ref{f:Dia1} and~\ref{f:Hus1}, and ${\beta_0'=1.7}$ in Figs.\,\ref{f:Dia2} and~\ref{f:Hus2}.
Both cases represent the first-order QPT between spherical and deformed ground-state shapes with the same critical value ${\lambda_{\rm c}=1}$.  
The phase-coexistence interval $(\lambda^*,\lambda^{**})$, demarcated by the spinodal and antispinodal points, coincides with ${(0.707,1.333)}$ for ${\beta_0'=\sqrt{2}}$ and with ${(0.460,1.257)}$ for ${\beta_0'=1.7}$.
As explained above, the U(5) dynamical symmetry is located at ${\lambda=0}$ for all $\beta_0'$ values, and the point ${\lambda=2}$ of the ${\beta_0'=\sqrt{2}}$ case is associated with the SU(3) dynamical symmetry.
Samples of the potential energy functions ${\cal V}^{\lambda}(x,y)$ along the ${y=0}$ line can be seen in the left columns of Figs.\,\ref{f:Hus1} and~\ref{f:Hus2}.

Figures~\ref{f:Dia1}--\ref{f:Hus2} depict not the level density itself, but its first derivative with respect to energy, $\frac{\partial}{\partial E}\overline{\rho}^{\lambda}(E)$. 
This is in agreement with the general classification of ESQPTs in ${f=2}$ systems, where we generically expect singularities in the first derivative of the level density and slope---see Sec.\,\ref{sec:cla}.
The smooth level density is calculated from the integral formula \eqref{claden} using the classical Hamiltonian forms \eqref{eq:H1cl} and \eqref{eq:H2cl}.
The integration is performed numerically in the whole above-defined finite phase space of the system.

The most complete information on the level density  is contained in Figs.\,\ref{f:Dia1} and \ref{f:Dia2}.
The top panels of these figures show $\frac{\partial}{\partial E}\overline{\rho}^{\lambda}(E)$ in the plane of control parameter ${\lambda\in[0,3.2]}$ and excitation energy ${E\in[0,3]}$ encoded in the shades of blue and red colors.
Blue color demarcates the domains of increasing level density (positive derivative), red indicates the domains of decreasing level density (negative derivative). 
The darker is the shade, the larger absolute value  takes the derivative at the corresponding place. 
White areas represent the domains where the level density is flat (has zero derivative).
In these domains, $\overline{\rho}^{\lambda}(E)$ is either equal to zero, which is the case beyond the finite energy domain $[E^{\lambda}_{\rm min},E^{\lambda}_{\rm max}]$ of the spectrum (i.e., below the blue and above red areas), or is fixed at a non-zero constant, which happens in some regions in the middle of the spectrum (between the blue and red areas).

The middle panels of both  Figs.\,\ref{f:Dia1} and \ref{f:Dia2} display critical curves where the level density derivative from the respective upper panel exhibits any kind of singularity.
These curves, which form the ESQPT critical borderlines in the $\lambda\times E$ plane, correspond to energies of various stationary points of classical dynamics.
The stationary points were determined by an exhaustive numerical procedure using the classical Hamiltonian forms \eqref{eq:H1cl} and \eqref{eq:H2cl}. 
Note that analytic determination was possible in some cases for ${\beta_0'=\sqrt{2}}$, see Ref.\,\cite{ref:Mace14}.

The dependencies of $\frac{\partial}{\partial E}\overline{\rho}^{\lambda}(E)$ on energy $E$ for four selected values of $\lambda$ are shown in the right columns of Figs.\,\ref{f:Hus1} and \ref{f:Hus2}.
The selected values correspond to the cuts (a)--(d) indicated in Figs.\,\ref{f:Dia1} and \ref{f:Dia2}.
In the level density plots shown in the respective panels (a)--(d) of Figs.\,\ref{f:Hus1} and \ref{f:Hus2} we observe upward and downward jumps as well as positive and negative divergences of the level density derivative.
These singularities appear precisely at the critical borderlines from the middle panels of Figs.\,\ref{f:Dia1} and~\ref{f:Dia2}.

Classification of  the ESQPT singularities in Figs.\,\ref{f:Dia1}--\ref{f:Hus2} is encoded in the types of the corresponding lines. 
The same line type is used consistently in the middle and bottom panels of  Figs.\,\ref{f:Dia1} and~\ref{f:Dia2}, as well as in Figs.\,\ref{f:Hus1} and~\ref{f:Hus2}.
The key is as follows:
\vspace{-\topsep}
\begin{itemize}
\setlength{\itemsep}{0pt}
\setlength{\parskip}{0pt}
\item[(i)] The full blue line corresponds to stationary points with index ${r=0}$ (minima of the Hamiltonian function). These cause upward jumps of the level density derivative.
\item[(ii)] The dashed blue lines indicate stationary points with index ${r=1}$ (saddles of the Hamiltonian function). They give rise to a positive logarithmic divergence of the level density derivative.
\item[(iii)] The dashed-dotted black line represents stationary points with index ${r=2}$ (saddles of the Hamiltonian function). These yield downward jumps of the level density derivative.
\item[(iv)] The dashed red line demarcates stationary points with index ${r=3}$ (saddles of the Hamiltonian function). Such points result in a negative logarithmic divergence of the level density derivative.  
\item[(v)] The full red line corresponds to stationary points with index ${r=4}$ (maxima of the Hamiltonian function). They cause upward jumps of the level density derivative, similar to type (i). 
\item[(vi)] The thick green lines are connected with the boundary $\partial{\bf\Omega}$ of the phase space. The corresponding singularities are divergences or jumps of the level density derivative.  
\end{itemize}
\vspace{-\topsep}

It turns out that most of the above level-density singularities follow from stationary points of the respective potential energy functions ${\cal V}^{\lambda}(x,y)$ (cf.\,the left panels of  Figs.\,\ref{f:Hus1} and \ref{f:Hus2}).
These form essentially the same types of ESQPTs as studied in parts~I and II \cite{ref:E1,ref:E2}.
However, some of the singularities result from the new features discussed in the present paper, either from the nontrivial kinetic energy term of the Hamiltonian (see Sec.\,\ref{sec:rol}), or from the boundedness of the compact phase space domain (see Sec.\,\ref{sec:bou}).

The \uvo{kinetic} type of ESQPTs can be recognized with the aid of the bottom panels of Figs.\,\ref{f:Dia1} and \ref{f:Dia2}, which show the coordinate $x$ and momentum $p_x$ corresponding to the stationary points responsible for the ESQPTs in the respective middle panels ($y$ and $p_y$ are not shown as they contain similar information).
Nonzero momentum values indicate the \uvo{kinetic} stationary points satisfying Eq.\,\eqref{stanotri} with ${\vecb{p}_0\neq 0}$.
This concerns a segment of the line (iv) in Fig.\,\ref{f:Dia1}, and a segment of line (iii) and whole lines (ii) and (iv) in Fig.\,\ref{f:Dia2}.

The \uvo{boundary} type of ESQPTs is connected with the critical borderlines in the middle panels of Figs.\,\ref{f:Dia1} and \ref{f:Dia2} which have no counterparts in the bottom panels.
The lines of type (vi) correspond to the overall minimum $E_{{\bf\partial\Omega}{\rm min}}^{\lambda}$ and maximum $E_{{\bf\partial\Omega}{\rm max}}^{\lambda}$ of the boundary energy function \eqref{E0}, which are independently of $\beta_0'$ given by the following simple analytic prescriptions:
\begin{equation}
(E_{{\bf\partial\Omega}{\rm min}}^{\lambda},E_{{\bf\partial\Omega}{\rm max}}^{\lambda})=\left\{\begin{array}{ll}
\left(1,1\I{+}\lambda^2\right) & {\rm for\ }\lambda\in[0,1),\\
\left(\frac{1+\lambda}{2},2\right) & {\rm for\ }\lambda\in[1,3),\\
\left(2,\frac{1+\lambda}{2}\right) & {\rm for\ }\lambda\in[3,\infty).
\end{array}\right.
\label{E0minmax}
\end{equation}
We note that the associated singularities of the level density (visible already in the first derivative, see Figs.\,\ref{f:Hus1} and \ref{f:Hus2}) are beyond the classification following from Eq.\,\eqref{expo} because of a high degree of degeneracy and non-separability of these points.
Additional singularities resulting from less degenerate stationary points of $E_{{\bf\partial\Omega}}^{\lambda}(\vecb{\phi})$ may be present in higher level density derivatives---their detection would be already beyond our numerical reach.

Comparing  the results for ${\beta_0'=\sqrt{2}}$ (Figs.\,\ref{f:Dia1} and \ref{f:Hus1}) and ${\beta_0'=1.7}$ (Figs.\,\ref{f:Dia2} and \ref{f:Hus2}), we see an overall similarity of phase diagrams, but note that the latter case exhibits a somewhat more complex system of ESQPTs.
The ${\beta_0'=1.7}$ energy surface generates a larger number of critical borderlines and a more sophisticated bifurcation pattern of stationary points.
In particular, we observe three ESQPTs associated with the \uvo{kinetic} type of stationary points, instead of just one such ESQPT present in the ${\beta_0'=\sqrt{2}}$ phase diagram.
Our numerical study of a wider range of parameter $\beta_0'$ showed that complexity of the phase diagram keeps increased within the whole interval ${\beta_0'\in(\sqrt{2},\sqrt{3})}$.
If $\beta_0'$ decreases below $\sqrt{2}$, the ESQPTs at $(\lambda,E)=(0,1)$ become degenerate (see Fig.\,\ref{f:Dia1}), which simplifies the picture at small values of $\lambda$. 
On the other hand, if $\beta_0'$ increases above $\sqrt{3}$, the spinodal point $\lambda^{*}$ disappears (implying that the spherical and deformed potential minima coexist at ${\lambda=0}$), which again leads to a less complicated structure of critical borderlines.
 
\subsubsection{Oscillatory level density}
\label{sec:Resosc}

\begin{figure}[t!]
\includegraphics[width=\linewidth]{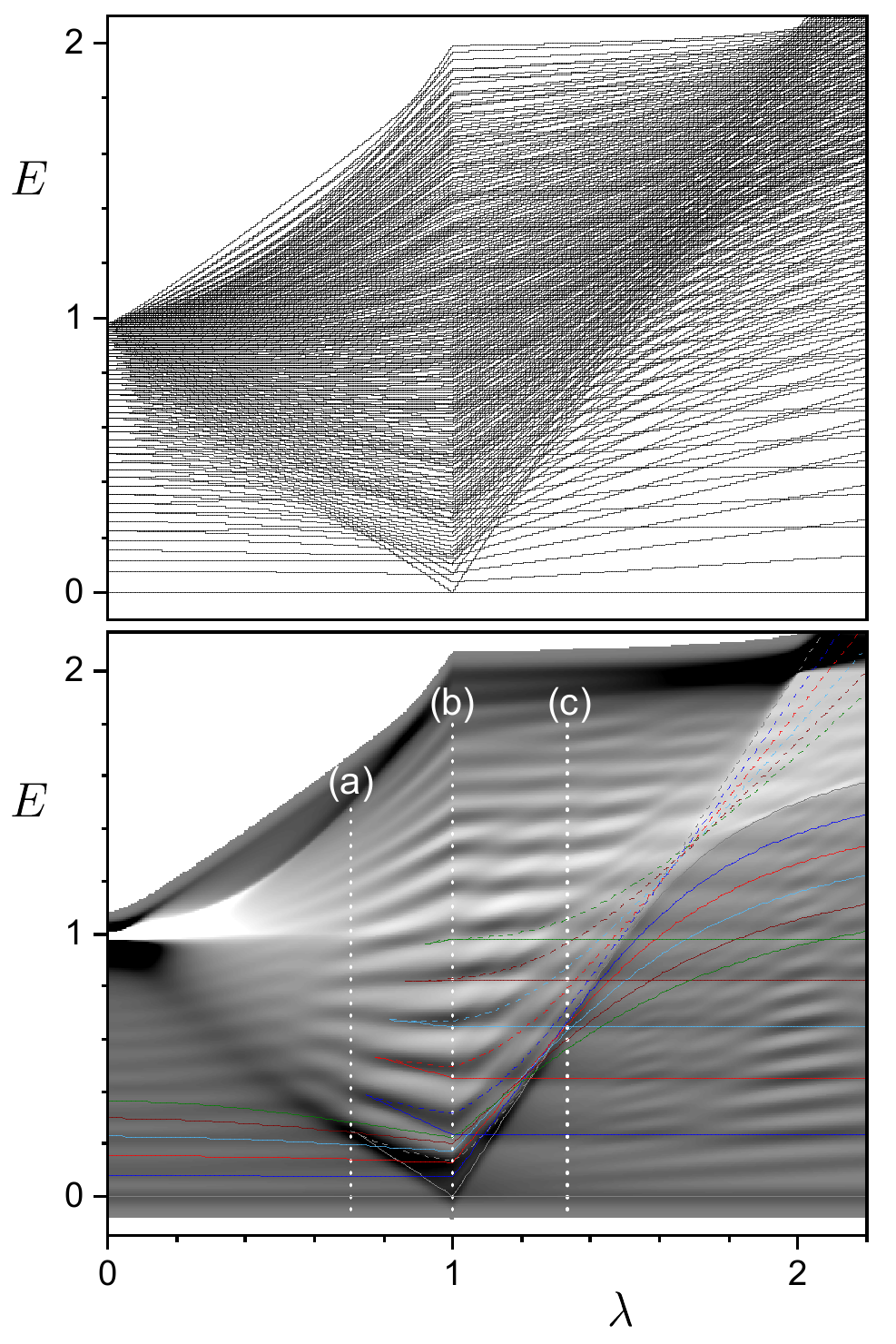}
\caption{
The spectrum of Hamiltonian \eqref{Hamib} with ${N=50}$ and ${\beta_0'=\sqrt{2}}$ as a function of parameter $\lambda$ (upper panel) and the corresponding oscillatory component $\widetilde{\rho}^{\lambda}(E)$ of the level density (lower panel).
The breaks of all levels (except the ${E_0^{\lambda}=0}$ ground state) in the upper panel at ${\lambda=\lambda_{\rm c}=1}$ result from the piecewise form of the Hamiltonian.
The size of $\widetilde{\rho}^{\lambda}(E)$  in the lower panel is expressed by the shades of gray (darker shades mark larger values).
Colored curves depict energies of stationary points of the excited potential energy surfaces (see text). 
The line types encode the numbers $N_\gamma$ of $\gamma$-excitation quanta consistently with Fig.\,\ref{f:Esur}.
Vertical lines (a), (b) and (c) demarcate position of the spinodal, critical and antispinodal points, respectively, where the potentials of Fig.\,\ref{f:Esur} are drawn.
The axes are dimensionless.}
\label{f:Osc}
\end{figure}

\begin{figure}[t!]
\includegraphics[width=0.8\linewidth]{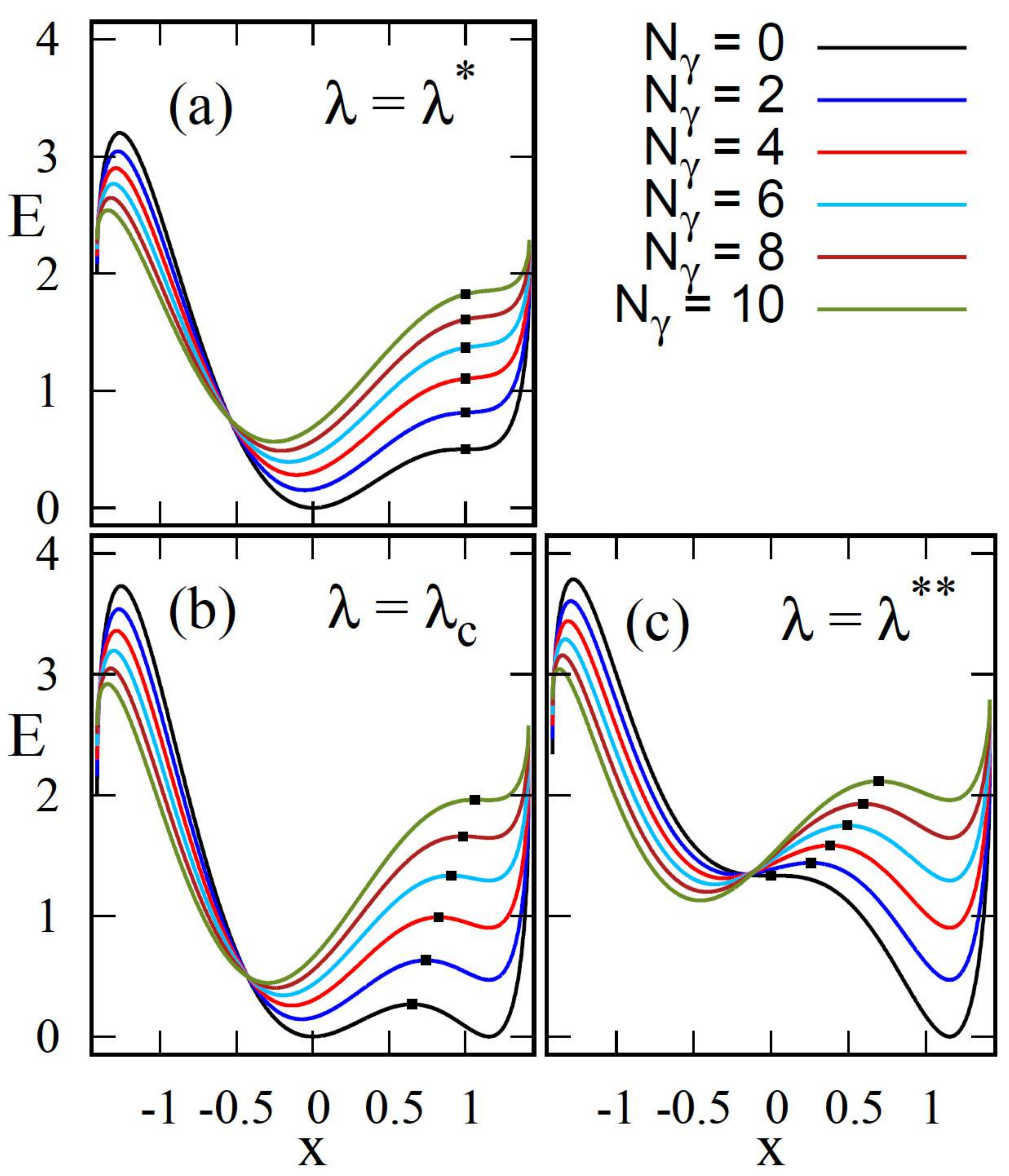}
\caption{Excited potential energy surfaces \eqref{Haprocoex} for three parameter values of Hamiltonian \eqref{Hamib} with ${\beta_0'=\sqrt{2}}$.
The condensate boson follows from Eq.\,\eqref{boco}, the excited mode is associated with the $\gamma$ boson \eqref{boga}.
We show the ${y=0}$ cuts of surfaces depending on ${(x,y)}$. 
Individual curves in each panel depict potentials with the given numbers of the $\gamma$ boson.
The values of $\lambda$ correspond to the spinodal (a), critical (b), and antispinodal (c) points of the ground-state potential (cf.\,Fig.\,\ref{f:Osc}).
For a more detailed explanation see Appendix~\ref{ApB}.
All axes are dimensionless.}
\label{f:Esur}
\end{figure}   

While the smooth component $\overline{\rho}^{\lambda}(E)$ of the level density captures the constitutive features that define the ESQPT of a given type in the infinite-size limit, the oscillatory component $\widetilde{\rho}^{\lambda}(E)$ carries information on finite-size effects.
The upper panel of Fig.\,\ref{f:Osc} displays the spectrum of ${L=0}$ eigenstates of the IBM Hamiltonian \eqref{Hamib} with ${N=50}$ for ${\beta_0'=\sqrt{2}}$ and ${\lambda\in[0,2.2]}$. 
This is a finite-$N$ realization of the semiclassical level density discussed in the previous paragraph, cf.\,Fig.\,\ref{f:Dia1}.
The lower panel of Fig.\,\ref{f:Osc} shows the oscillatory component of the level density derived from the spectrum in the upper panel.
It was calculated from the formula
\begin{equation}
\widetilde{\rho}^{\lambda}(E)=\sum_i\delta_{\sigma}(E-E_i^{\lambda})-\overline{\rho}^{\lambda}(E)\,,
\label{oden}
\end{equation}
where $\delta_{\sigma}(x)$ stands for a Gaussian of a width $\sigma$ smaller than a typical spacing between energy levels $E_i^{\lambda}$, so ${\sigma<1/\overline{\rho}^{\lambda}(E)}$, and the smooth component $\overline{\rho}^{\lambda}(E)$ is evaluated from the semiclassical integral in Eq.\,\eqref{claden}.
We observe a regular alternation of denser and sparser parts of the spectrum, indicated by darker and lighter shades, respectively.

It turns out that the structures in Fig.\,\ref{f:Osc} exhibit the separability-enhanced precursors of ESQPTs  discussed in Sec.\,\ref{sec:sep}.
In part~II of this series \cite{ref:E2} we showed, using a simpler model with partially separated $x$ and $y$ degrees of freedom, that stationary points of an effective potential for $x$ (obtained by fixing the averages of variables related to the $y$ degree of freedom) give rise to precursors of ESQPT structures in spectra of excitations in $x$ direction built on individual excitations in $y$  direction.
The effective number of degrees of freedom associated with these structures was ${f_{\rm eff}=1}$ rather than ${f=2}$.

A similar effect can be demonstrated in the present model, in which the two partially separated degrees of freedom are the $\beta$ and $\gamma$ vibrational modes.
For the ${\beta_0'=\sqrt{2}}$ Hamiltonian \eqref{Hamib} it turns out that the energy ratio between the $\beta$ and $\gamma$ phonon energies is ${\varepsilon_{\beta}/\varepsilon_{\gamma}=(2\lambda-1)/3}$ \cite{ref:Mace14}. 
So the $\gamma$ mode is markedly faster than the $\beta$ mode for $\lambda\lesssim 1.5$, and in this parameter domain we expect bands of $\beta$-vibrational states built on individual $\gamma$-vibrational states in the spectrum.
Hence the formula \eqref{Haprocoex} can be applied with the condensate boson $\hat{B}^{\dag}(\tilde{\alpha})$ from Eq.\,\eqref{boco} and the orthogonal excitation boson $\hat{B}^{\dag}(\tilde{\alpha}_{\perp})$ associated with $\hat{B}^{\dag}_{\gamma}(\gamma)$ from Eq.\,\eqref{boga}.
As the coefficients in these equations have no imaginary components (no momenta), the excited energy surfaces resulting from this construction represent surfaces of potential energy depending on coordinates $\beta$ and $\gamma$, or equivalently $x$ and $y$.

Excited potentials (their ${y=0}$ cuts) for three choices of parameter $\lambda$ and several numbers $N_{\gamma}$ of the $\gamma$-vibrational quanta are shown in Fig.\,\ref{f:Esur} (the reason why $N_{\gamma}$ takes only even values is explained in Appendix~\ref{ApB}). 
The values of $\lambda$ are chosen at the spinodal point ${\lambda^*=\frac{1}{\sqrt{2}}}$ (the point, where the ground-state potential develops a secondary minimum corresponding to a deformed shape), critical point ${\lambda_{\rm c}=1}$ (where both minima of the ground-state potential become degenerate), and the antispinodal point ${\lambda^{**}=\frac{4}{3}}$ (where the minimum associated with the spherical shape disappears).
We observe that with increasing $N_{\gamma}$ the minimum corresponding to the spherical shape becomes deeper in comparison with the minimum associated with the deformed shape.
As a consequence, the phase evolution of excited states is retarded with respect to the ground state (the critical value of parameter $\lambda$ at which both minima of the excited potential get degenerate increases above the ground-state critical point $\lambda_{\rm c}$).
Although we do not evaluate the dependence of the excited potential energy surfaces on the $y$ variable, we assume that for lower excitation energies the ${y=0}$ cut represents the most substantial information due to expected averaging of excited dynamics at ${\ave{\gamma}=0}$.

Energies of the lowest stationary points of the excited potentials from Fig.\,\ref{f:Esur} are displayed by the colored curves in the lower panel of Fig.\,\ref{f:Osc}  for the whole interval ${\lambda\in[0,2.2]}$.
Each line type corresponds to the same number $N_{\gamma}$ of the $\gamma$-excitation quanta (the color code is the same as in Fig.\,\ref{f:Esur}). 
The three curves of the same line type with increasing energy represent, consecutively, the primary minimum, secondary minimum and intermediate maximum (probably a saddle point of the whole dependence on $x$ and $y$) of the respective excited potential.

We see that at lower excitation energies there is a rather strong correlation of the colored curves with the oscillatory structures in Fig.\,\ref{f:Osc}.
This illustrates the above-discussed finite-size effects induced by partial separability.
In general, minima of an effective ${f_{\rm eff}=1}$ potential lead to a step-like increases of the level density while a local maximum indicates a peak in the level density.
Such structures can be indeed distinguished in Fig.\,\ref{f:Esur}.
We clearly observe their gradual shift to higher parameter values, which agrees with the above-discussed evolution of excited potentials.
For higher energies, the correspondence gradually deteriorates.
This indicates that the large-$N$ approximation involved in the foundations of the present generalized coherent state method becomes worse as the ratio $N_{\gamma}/N$ increases.
Moreover, for highly excited potentials the whole ${(x,y)}$ dependence may become increasingly relevant.

We point out that the spectra of our model contain, besides the structures in $\widetilde{\rho}^{\lambda}(E)$ related to partial separability, also some additional oscillatory structures with a different origin.
These can be noticed in both panels of Fig.\,\ref{f:Osc}  in the domain ${\lambda>1}$ for energies ${E<\lambda-1}$, mostly close to the values of $\lambda$ for which the energy ratio ${\varepsilon_{\beta}/\varepsilon_{\gamma}}$ becomes a simple rational number (particularly ${1,\frac{2}{3},\frac{1}{2},\frac{2}{5},\frac{1}{3}}$).
It was shown in Ref.\,\cite{ref:Mace14} (cf.\,Fig.\,14 therein) that these clustering effects are related to splitting of classical phase-space tori near the above resonances via a mechanism following from the classical chaos theory.

\subsubsection{Links to previous results}
\label{sec:Redis}

To conclude the present ESQPT case study, let us briefly address the relation of our results to those of Ref.\,\cite{ref:Zhan16}. 
The cited work applied the simplified IBM Hamiltonian \eqref{Haha} along several symmetry-connecting paths in parameters $\eta$ and $\chi$.
It turned out that energy dependencies of expectation values of the $d$-boson number ${\hat{n}_d}$ in individual excited states with ${L=0}$ exhibit some special structures near the ESQPT borderlines, considered in Ref.\,\cite{ref:Zhan16} only for cases of the potential maximum related to spherical deformation and the saddle point related to oblate deformation.

These structures can be justified in terms of the theoretical background outlined in Sec.\,\ref{sec:cla}.
In particular, we saw that the ESQPT singularities of the level density are accompanied by the same types of singularities in the level flow, i.e., for ${f=2}$ systems generically by nonanalyticities of the smoothed slope of the spectrum.
For the Hamiltonian \eqref{Haha} it follows from Eq.\,\eqref{hell} that
\begin{eqnarray}
-\frac{\partial E_i^{(\eta,\chi)}}{\partial\eta}&\I{=}&\matr{\psi_i^{(\eta,\chi)}}{\hat{n}_d}{\psi_i^{(\eta,\chi)}}\I{+}\frac{\matr{\psi_i^{(\eta,\chi)}}{(\hat{Q}^{\chi}\I{\cdot}\hat{Q}^{\chi})}{\psi_i^{(\eta,\chi)}}}{4N}\,,
\nonumber\\
-\frac{\partial E_i^{(\eta,\chi)}}{\partial\chi}&\I{=}&\eta\frac{\matr{\psi_i^{(\eta,\chi)}}{\frac{\partial}{\partial\chi}(\hat{Q}^{\chi}\I{\cdot}\hat{Q}^{\chi})}{\psi_i^{(\eta,\chi)}}}{4N}\,,
\label{hellochi}
\end{eqnarray}
where $\ket{\psi_i^{(\eta,\chi)}}$ is the $i$th eigenstate with energy $E_i^{(\eta,\chi)}$ at the given point $(\eta,\chi)$.
Since energy derivatives on the left-hand side of these formulas (if averaged out over the ${L=0}$ eigenstates within a narrow interval around the given energy) exhibit precursors of the ESQPT singularities, the same must be true for the expectation values on the right-hand side.
The latter involve the ${\hat{n}_d}$ expectation value, either explicitly as in the first formula, or via expansions of operators $(\hat{Q}^{\chi}\I{\cdot}\hat{Q}^{\chi})$ and $\frac{\partial}{\partial\chi}(\hat{Q}^{\chi}\I{\cdot}\hat{Q}^{\chi})$ into the Casimir invariants of all subalgebras of U(6); cf. formula (5) and Table~I in Ref.\,\cite{ref:Cejn98}.
Hence the  locally averaged $\hat{n}_d$ expectation values naturally show some anomalous energy dependencies near the ESQPT critical energies.

It shall be pointed out that anomalous behavior at ESQPT energies may be anticipated for locally smoothed expectation values of almost any generic observable $\hat{A}$.
Indeed (returning to a general Hamiltonian $\hat{H}^{\lambda}$), since the slope ${\frac{\partial}{\partial\lambda}E_i^{\lambda}}$ of individual levels can be decomposed into a sum of terms ${\matr{\psi_i^{\lambda}}{\hat{A}}{\psi_i^{\lambda}}}$ and ${\matr{\psi_i^{\lambda}}{\frac{\partial}{\partial\lambda}\hat{H}^{\lambda}-\hat{A}}{\psi_i^{\lambda}}}$, it is natural to assume that the anomaly in the smoothed slope is shared by both these terms.
Note that instead of spectral averages of expectation values one can consider the whole Peres lattices ${E_i^{\lambda}\times\matr{\psi_i^{\lambda}}{\hat{A}}{\psi_i^{\lambda}}}$ constructed for all Hamiltonian eigenstates.
Examples (from another model) of ESQPT anomalies simultaneously reflected by Peres lattices of several observables can be found in Re.\,\cite{ref:Kloc17a}.

The present analysis of ESQPT effects in the IBM differs from one of Ref.\,\cite{ref:Zhan16} in several aspects.
First, we focus on the description of ESQPTs in terms of the level density instead of expectation values.
Peres lattices of observable ${\propto\sqrt{\hat{n}_d}}$ for the present Hamiltonian were studied in Ref.\,\cite{ref:Mace14}.
Second, our aim was to localize and classify spectral effects of all singularities in classical dynamics and to anchor them in the general theory developed in parts~I and II \cite{ref:E1,ref:E2} and Sec.\,\ref{sec:MB}. 
Using this exhaustive treatment, we demonstrate that for a genuine ${f=2}$ many-body system the singularities appear in large numbers and are of rather diverse types.
Third, we used the more complex IBM Hamiltonian from Eq.\,\eqref{Hamib}, which describes the first-order transition between spherical and deformed nuclear shapes with flexible properties.
This allowed us to continue our previous study of ESQPT structures accompanying a generic first-order ground-state QPT.

Note that Ref.\,\cite{ref:Zhan17} extends the study of Ref.\,\cite{ref:Zhan16} to states with ${L>0}$.
However, relaxing the angular momentum constraint, one generally leaves the ${f=2}$ subset of IBM dynamics.
So the ${L>0}$ results are beyond the reach of the present approach.
In particular, for ${f>2}$ systems one generally expects ESQPT singularities in higher derivatives of the level density and flow.

\section{Summary and outlook}
\label{sec:Sum}

This paper is part~III of a series devoted to the study of ESQPTs in bound quantum systems with ${f=2}$ effective degrees of freedom.
In parts~I and II \cite{ref:E1,ref:E2}, we considered systems with standard Hamiltonians of the type ${\hat{H}^{\lambda}=\frac{1}{2}\hat{\vecb{p}}^2+V^{\lambda}(\hat{\vecb{q}})}$.
Here we have extended our view to genuine quantum many-body systems with Hamiltonians written in terms of particle creation and annihilation operators. 
In particular, we have studied bound systems composed of several mutually interacting bosonic species, assuming that the total number of particles is conserved.  
The analysis has been first performed for a general system of the above type, and then illustrated by a concrete example of the interacting boson model of nuclear collective dynamics.

The present work can be seen as a realistic implementation of the results discussed in parts~I and II \cite{ref:E1,ref:E2}, but also as a study of new ESQPT-related effects that emerge in the many-body context.
In particular, we have discussed additional singularities of the level density that appear in connection with (i) coordinate-dependent kinetic terms of a generic many-body Hamiltonian and (ii) finiteness of the  phase space resulting from the conserved number of particles.
We have also shown that finite-size oscillatory structures in the level density, caused by partial separability of dynamics (as thoroughly studied in part~II \cite{ref:E2}), can be effectively described  by excited energy surfaces resulting from the formalism of generalized coherent states.
These structures, which originate in dynamics reduced to a single degree of freedom, can in some cases and for moderate values of $N$ overwhelm the precursors of the ${N\to\infty}$ ESQPT signatures in the smooth component of the level density.

It should be pointed out that one may hardly hope in a direct experimental evidence of the effects discussed here in collective dynamics of nuclei at low energies, where the IBM is applicable.
The density of states is too low in these domains to provide an unambiguous signature of an ESQPT, while at higher energies with larger state densities, the number of effective degrees of freedom grows very rapidly in nuclei.
On the other hand, the nuclear IBM offers a powerful tool to verify the conclusions of our general analysis, which may turn relevant in other contexts of nuclear physics.

The present series of works can be naturally continued by investigating other types of quantum many-body systems with two effective degrees of freedom.
First, the analysis can be done for some bound fermionic systems.
If the  ${f=2}$ collective dynamics of such a system is described by means of a finite dynamical algebra, there might be a close link to the present work enabled by a suitable bosonic representation of the algebra \cite{ref:Klei91}.
Second, a similar study as presented here can be performed in the context of extended lattice systems with collective excitations, like two-dimensional crystals \cite{ref:Diet13,ref:Iach15,ref:Diet17}.

\section*{Acknowledgments}
\label{sec:Ack}
M.M. thanks F.\,Iachello for insightful discussions and acknowledges support by the Czech MEYS project RVO:68081731.
P.S. and P.C. acknowledge funding by the Charles University Project UNCE/SCI/013.
A.L. appreciates support by the Israel Science Foundation.

\appendix

\section{Generalized projective coherent states}
\label{ApB}

As an extension of Eq.\,\eqref{proco}, we define a generalized coherent state given by the following prescription:
\begin{eqnarray}
\left|\tilde{\alpha},\{N_l\}\right\rangle=\frac{1}{\sqrt{N_0!N_1!\cdots N_m!}}\times
\qquad\qquad\qquad\quad
\nonumber\\
\hat{B}^{\dag}(\tilde{\alpha})^{N_0}\hat{B}_{\perp 1}^{\dag}(\tilde{\alpha})^{N_1}\cdots\hat{B}_{\perp m}^{\dag}(\tilde{\alpha})^{N_m}\ket{0}.
\qquad
\label{procoex}
\end{eqnarray}
Here, $\hat{B}^{\dag}(\tilde{\alpha})$ is a boson creation operator from  Eq.\,\eqref{co}, with ${\{\tilde{\alpha}\equiv\{\alpha_k/\lVert\alpha\rVert\}_{k=1}^n}$ denoting normalized expansion coefficients, while operators ${\hat{B}_{\perp l}^{\dag}(\tilde{\alpha})\equiv\hat{B}^{\dag}(\tilde{\alpha}_{\perp l})}$ with ${l=1,...,m}$ create normalized bosons in states perpendicular to $\hat{B}^{\dag}(\tilde{\alpha})$ and to each other (expansion coefficients $\tilde{\alpha}_{\perp l}$ satisfy the orthonormality constraints ${\tilde{\alpha}_{\perp l}\cdot\tilde{\alpha}=0}$ and ${\tilde{\alpha}_{\perp l}\cdot\tilde{\alpha}_{\perp l'}=\delta_{ll'}}$, and therefore depend on $\tilde{\alpha}$).
Powers $N_l$ represent the numbers of bosons in these states.
As the total number of bosons $N$ is fixed, we can write ${N_0=N-N_1-N_2-\cdots-N_m}$. 

The states \eqref{procoex} may be used in an extended variational procedure \cite{ref:Capr05} to approximate the eigensolutions of an interacting boson Hamiltonian $\hat{H}$ conserving the total particle number.
The ground state for large $N$ is captured by a condensate $\propto\hat{B}^{\dag}(\tilde{\alpha})^N\ket{0}$ from Eq.\,\eqref{proco}, that is by the state \eqref{procoex} with ${\{N_0,N_1,...,N_m\}=\{N,0,...,0\}}$.
The optimal set of parameters $\tilde{\alpha}$ results from the minimization of the energy surface \eqref{Haproco}.
Excited states are then described with the aid of the supplementary bosons $\hat{B}^{\dag}_{\perp l}(\tilde{\alpha})$, which depend via the orthogonality constraint on the ground-state parameters $\tilde{\alpha}$. 
These bosons must be selected so that they represent the proper excitation modes of the system.
By increasing the ${l>0}$ occupation numbers $\{N_l\}$ in Eq.\,\eqref{procoex}, we obtain the bands of excited states of various types.

In analogy to the use of condensate states \eqref{proco} for the construction of classical Hamiltonian function ${\cal H}(\tilde{\alpha},N)$ in the phase space, see Eq.\,\eqref{Haproco}, the generalized coherent states \eqref{procoex} allow one to define classical energy surfaces corresponding to individual excited states.
They are given by expressions
\begin{equation}
{\cal H}_{\rm ex}(\tilde{\alpha},\{N_l\})=\frac{1}{N}\matr{\tilde{\alpha},\{N_l\}}{\hat{H}}{\tilde{\alpha},\{N_l\}},
\label{expos}
\end{equation}
which prescribe specific dependences of energy on coordinates and momenta (encoded in variables $\tilde{\alpha}$) for the states with given numbers $\{N_l\}$ of elementary excitations.
Let us stress that the surfaces ${\cal H}_{\rm ex}$ are not designed for a variational determination of excited states as those would not in general be orthogonal to the variational ground state and to each other.
However, it turns out (see Sec.\,\ref{sec:Resosc}) that stationary points of these surfaces have a strong impact on the oscillatory structures of the level density.

Applying these ideas in the framework of the IBM, we first define the ground-state condensate boson: 
\begin{equation}
\hat{B}^{\dag}(\beta,\gamma)=\underbrace{\sqrt{1\I{-}\tfrac{\beta^2}{2}}}_{s(\beta)}\hat{s}^{\dag}+\frac{\beta\cos\gamma}{\sqrt{2}}\hat{d}_{0}^{\dag}+\frac{\beta\sin\gamma}{2}(\hat{d}_{-2}^{\dag}\I{+}\hat{d}_{+2}^{\dag}).
\label{boco}
\end{equation}
Note that variables ${\beta\in[0,\sqrt{2}]}$ and ${\gamma\in[0,2\pi)}$ correspond to the standard nuclear shape parameters written in the finite-range convention \cite{ref:Klei81}. 
The components with $\hat{d}_{\pm 1}^{\dag}$ bosons vanish due to the transformation to the intrinsic frame \cite{ref:IBMbook}.
Making associations ${\hat{b}^{\dag}_0\equiv\hat{s}^{\dag}}$ and $\{\hat{b}^{\dag}_1, ... ,\hat{b}^{\dag}_5\}\equiv\{\hat{d}_{-2}^{\dag}, ... ,\hat{d}_{+2}^{\dag}\}$, the expansion coefficients in Eq.\,\eqref{proco} read as follows: 
$\tilde{\alpha}=\left(s(\beta),\frac{1}{2}\beta\sin\gamma,0,\frac{1}{\sqrt{2}}\beta\cos\gamma,0,\frac{1}{2}\beta\sin\gamma\right)$.
Reality of these coefficients indicates that the function ${\cal H}(\tilde{\alpha},N)$ in Eq.\,\eqref{Haproco} is not the full Hamiltonian, but only the potential energy depending on shape parameters $\beta,\gamma$.

In addition to the condensate boson \eqref{boco}, we introduce bosons associated with $\beta$ and $\gamma$ vibrational modes \cite{ref:Levi87}: 
\begin{eqnarray}
\hat{B}^{\dag}_{\beta}(\beta,\gamma)&\!\!=\!\!&-\frac{\beta}{\sqrt{2}}\hat{s}^{\dag}\I{+}s(\beta)\cos\gamma\,\hat{d}_{0}^{\dag}\I{+}\frac{s(\beta)\sin\gamma}{\sqrt{2}}(\hat{d}_{-2}^{\dag}\I{+}\hat{d}_{+2}^{\dag}),
\nonumber\\
\hat{B}^{\dag}_{\gamma}(\gamma)&\!\!=\!\!&-\sin\gamma\,\hat{d}_{0}^{\dag}\I{+}\frac{\cos\gamma}{\sqrt{2}}(\hat{d}_{-2}^{\dag}\I{+}\hat{d}_{+2}^{\dag}).
\label{boga}
\end{eqnarray}
The single-particle states associated with $\hat{B}^{\dag}(\beta,\gamma)$, $\hat{B}^{\dag}_{\beta}(\beta,\gamma)$ and $\hat{B}^{\dag}_{\gamma}(\gamma)$ are normalized and mutually orthogonal.
Note that one can introduce yet three other collective bosons, namely ${\tfrac{1}{\sqrt{2}} (\hat{d}^\dag_{+1}\pm\hat{d}^\dag_{-1})}$ and ${\tfrac{1}{\sqrt{2}} (\hat{d}^\dag_{+2}-\hat{d}^\dag_{-2})\equiv\hat{B}^\dag_z}$, which together with the above bosons  form a complete set replacing the original six bosons $\hat{s}^{\dag}$ and $\{\hat{d}_m^{\dag}\}$.
In the large $N$ limit, and in absence of an axial symmetry of the intrinsic shape (i.e., for $\gamma$ not equal to integer multiples of $\frac{\pi}{3}$), the additional bosons characterize collective rotations, while bosons \eqref{boco} and \eqref{boga} approximate intrinsic excitations.
In particular, multiple $\gamma$ excitations can be written as ${\propto\hat{B}^{\dag}_{\gamma}(\gamma)^{N_\gamma}\hat{B}^{\dag}(\beta,\gamma)^{N-N_\gamma}\ket{0}}$ with ${N_{\gamma}=1,2,3,\dots}$. 
If the intrinsic shape is axially symmetric around $z$, i.e., ${\gamma = 0}$ or $\pi$, the ${\hat{B}^{\dag}_{\gamma}(\gamma)}$ excitation can be combined with $\hat{B}^\dag_z$ in such a way that the projection $K$ of angular momentum on the symmetry axis is conserved. 
The multiple $\gamma$ excitations with ${K=0}$  exist only for $N_{\gamma}$ even and are given by $\propto(\hat{d}^\dag_{+2}\hat{d}^\dag_{-2})^{N_\gamma/2}\hat{B}^\dag(\beta,\gamma)^{N-N_\gamma}\ket{0}$. 
It can be shown, however, that the latter form of $\gamma$ excitations yields the same $\beta$ dependeces of excited potentials as the ${\gamma = 0}$ or $\pi$ cuts of the potentials obtained with the aid of ${\hat{B}^{\dag}_{\gamma}(\gamma)}$.
This is relevant for the explanation of Fig.\,\ref{f:Osc}.

\end{document}